\documentclass[%
amsmath,
reprint,
superscriptaddress,
prc,
]{revtex4-1}

\usepackage[T1]{fontenc}
\usepackage{textcomp}
\usepackage{txfonts}
\usepackage{bm}
\usepackage{graphicx}
\usepackage{dcolumn}
\usepackage{xcolor}
\usepackage{lipsum} 
\usepackage{mathtools}
\usepackage{sidecap}
\usepackage{booktabs}
\usepackage{float}
\usepackage{ragged2e}
\usepackage{anyfontsize}
\usepackage{braket}
\usepackage{hyperref}

\usepackage{chngcntr}

\hypersetup{colorlinks=true, linkcolor=blue, citecolor=blue, urlcolor=blue}

\begin{document}

\title{Microscopic triaxial quadrupole-octupole collective Hamiltonian for low-energy nuclear excitations}

\author{J. Xiang}
\affiliation{College of Physics and Electronic Engineering, Chongqing Normal University, Chongqing 401331, China}
\author{J. Zhao}
\affiliation{Center for Circuits and Systems, Pengcheng Laboratory, Shenzhen, China, 518055}

\author{Z. P. Li}
\email{zpliphy@swu.edu.cn}
\affiliation{School of Physical Science and Technology, Southwest University, Chongqing 400715, China}


\author{D. Vretenar}
\affiliation{Physics Department, Faculty of Science, University of Zagreb, Croatia}

\date{\today}

\begin{abstract}
We present a microscopic triaxial quadrupole-octupole collective Hamiltonian (TQOCH) that unifies collective rotations, quadrupole-octupole vibrations, and their couplings to model low-lying nuclear states of both parities. The TQOCH's dynamics are governed by collective parameters derived from multidimensionally constrained covariant density functional theory. The model's validity is demonstrated through calculations of $^{152}$Sm, including its deformation energy surfaces, excitation spectra, and transition probabilities. As a predictive tool, the TQOCH probes complex nuclear phenomena like shape coexistence and phase transitions, directly connecting their microscopic origins to spectroscopic observables.
\end{abstract}

\maketitle


\section{\label{sec:I}Introduction}
Nuclear shape remains a cornerstone of nuclear physics, with its understanding continuously deepened by experimental and theoretical progress. Key experimental signatures include rotational spectra, electromagnetic transition rates, magnetic moments, and the characteristics of K isomers \cite{RingBook,Bohrbook1975,DracoulisRPP2016,HerzbergNature2006}.
While quadrupole deformation (non-zero $\beta_{20}$) is well-established-evidenced by the I(I+1) energy pattern of a symmetric rigid rotor—triaxial deformation (non-zero $\beta_{20}$ and $\beta_{22}$) introduces greater complexity \cite{MollerPRL2006,LiPRC2010f,LuPRC2012,FuPRC2013,LiJPG2016,YangPRC2021}. Phenomena such as low-lying $K^\pi=2^+$ bands with specific staggering \cite{DavydovNP1958,SevrinPRC1987,ZamfirPLB1991,McCutchanPRC2007}, low-spin signature inversion \cite{HamamotoPLB1990,RiedingerPPNP1997251}, chiral doublet bands \cite{FrauendorfNPA1997,MengJPG2010,WangCPC2020,ChenNPN2020}, and wobbling bands \cite{HamamotoPRC2002,HamamotoPRC2003,Petrache2024chirality,ChenQB2016PRC,ChenQB2014PRC} are all considered key indicators of triaxiality in various nuclides. 

The study of reflection-asymmetric octupole deformation ($\beta_{30}$), first observed in the 1950s \cite{AsaroPR1953,StephensPR1955}, has advanced considerably with the development of rare isotope facilities. Research into axially symmetric octupole deformation has now expanded to multiple mass regions, including $A\sim70$~\cite{HanIJMPE2023,AnujPS2023,XiaoPRC2022,ZhangPRL2022,XuPLB2022,WangPRC2022,MukherjeePRC2022,RajbanshiPRC2021,BhattacharyaPRC2019,GregorEPJA2017}, $(Z\sim56, N\sim60)$~\cite{PandeyCPC2023,LvPRC2022,ChenPRC2016,TestovPRC2021}, $(Z\sim56, N\sim88)$~\cite{BrewerNPA2023,YagiPRC2022,GuoPLB2020,PetrachePRC2020,MorsePRC2020,MajolaPRC2019,BhattacharyyaPRC2018,BucherPRL2017,WangEPJA2017,HuangPRC2016}, and $(Z\sim88, N\sim136)$~\cite{ButlerPS2024,Rey-hermePRC2023,SpagnolettiPRC2022,YadavPRC2022,ButlerPRL2020,BerryPRC2020,RaletPLB2019,VerstraelenPRC2019,BarzakhPRC2019,PragatiPRC2018,SpiekerPRC2018,ParrPRC2016,HensleyPRC2017,machEPJA2016}. Remarkably, evidence for octupole deformation has been found even in the double-magic nucleus $^{208}$Pb, indicated by its substantial octupole moment \cite{HendersonPRL2025}.

Tetrahedral deformation ($\beta_{32}$), a type of axial-breaking octupole deformation, has garnered significant interest for its potential to create nuclei with tetrahedral shapes \cite{TakamiPLB1998,DudekPRL2002,SchunckPRC2004,DudekPRL2006,ChenPRC2008}. Its role has been explored with diverse theoretical approaches, such as macroscopic-microscopic models \cite{DudekPRL2002,DudekIJMPE2007,DudekPRL2006,SchunckPRC2004,DudekPS2014,JachimowiczPRC2017,YangPRC2022a,YangPRC2022b,SkalskiPRC1991}, Skyrme-Hartree-Fock-Bogoliubov theory~\cite{OLBRATOWSKIIJMPE2006,YamagamiNPA2001,TakamiPLB1998}, reflection asymmetric shell models~\cite{chenNPA2010,GaoCPL2004}, algebraic cluster models~\cite{BijkerPRL2014}, ab-initio calculations within nuclear lattice effective field theory~\cite{EpelbaumPRL2014}, 3D lattice space solutions of covariant density functional theory (CDFT)~\cite{XuPRC2024}, and multidimensionally constrained CDFT~\cite{ZhaoPRC2017,ZhaoPRC2012a,ZhaoPRC2024}. Related shape vibrations have also been studied using models like the interacting boson model (IBM)~\cite{EngelPRL1985,EngelNPA1987}, quasiparticle random phase approximation based on Hartree-Fock-BCS~\cite{LocPRC2023}, and the generator coordinate method~\cite{SkalskiNPA1993,ZbereckiPRC2006,ZbereckiPRC2009}. 

Experimentally, the search for tetrahedral symmetry has been challenging; early investigations (2010-2017) in nuclei like  $^{108}$Zr~\cite{SumikamaPRL2011}, $^{156}$Gd~\cite{BarkPRL2010,DoanPRC2010,DudekPRL2006,JentschelPRL2010},  $^{156}$Dy~\cite{HartleyPRC2017}, and $^{230,232}$U~\cite{NtshangasePRC2010} found no evidence. A recent breakthrough by S. Basak {\it et al.} claims the first experimental evidence—a tetrahedral band with a $3^-$ bandhead at 1933.5 keV in $^{152}$Sm \cite{BasakPRC2025}. Nevertheless, this finding remains incomplete as the crucial reduced transition strengths for this band are still missing.

Consequently, confirming the discovery of tetrahedral deformation demands a concerted effort. Experimentally, this involves comprehensive measurements of low-lying positive- and negative-parity bands and their electromagnetic transitions. Theoretically, a microscopic model is required that self-consistently includes non-axial quadrupole and octupole deformations to reproduce this data, thereby identifying band shapes and configurations to confirm the tetrahedral structure \cite{DudekPRL2006}. A step in this direction is the quadrupole-octupole collective Hamiltonian (QOCH) by Dobrowolski \textit{et al.}, which uses a macroscopic-microscopic approach in a nine-dimensional collective space \cite{DobrowolskiPRC2016,DobrowolskiPRC2018}. However, a key shortcoming of this model is its lack of coupling between quadrupole and octupole degrees of freedom. This omission renders it unable to describe secondary minima and their bands, a critical flaw given that these very couplings are known to induce higher-order octupole deformations \cite{ButlerRMP1996,DudekPRL2006}.

Established over the last decade, the multidimensionally constrained covariant density functional theory (MDC-CDFT) self-consistently includes axially-asymmetric and reflection-asymmetric deformations ($\beta_{20},\beta_{22},\beta_{30},\beta_{32}$)~\cite{LuPRC2012}. Its successful applications are diverse~\cite{ZhouPS2016}, spanning the calculation of fission barriers in actinides~\cite{LuEPJ2012,LuPRC2014,LuPS2014,LuPRC2012,ZhaoPRC2015} and superheavy nuclei~\cite{MengSCPMA2020}, investigation of higher-order deformations in superheavy nuclei~\cite{WangCPC2022}, non-axial octupole $Y_{32}$ correlations in $N=150$ isotones~\cite{ZhaoJPRC2012} and Zr isotopes~\cite{ZhaoPRC2017}, axial octupole $Y_{30}$ correlations in $M\chi{D}$~\cite{LiuCPRL2016}, $\alpha$ and cluster decays~\cite{ZhaoJPRC2023}, shape evolution~\cite{LuPRC2011,LuPRC2014}, and fission dynamics~\cite{ZhaoJPRC2015,ZhaoJPRC2016,ZhaoJPRC2019,ZhaoJPRC2019b,ZhaoJPRC2020,ZhaoJPRC2022,ZhaoJPRC2022b}.

The multidimensionally constrained CDFT has been extended to incorporate angular momentum and parity projections, restoring broken rotational and parity symmetries under $V_4$ symmetry~\cite{WangCTP2022}, and was successfully applied to describe the structure of $^{96}$Zr~\cite{RongPLB2023}. However, a comprehensive description of the low-lying positive- and negative-parity bands$-$which arise from vibrations, rotations, and their couplings$-$as well as the mechanism of shape coexistence, requires the inclusion of shape fluctuations.

In the present work, we develop a microscopic collective Hamiltonian based on the multidimensionally constrained CDFT that includes axial and triaxial quadrupole-octupole shape vibrations, rotations, and their couplings. This framework incorporates quadrupole-octupole couplings at both the static mean-field and dynamical levels, enabling a fully microscopic description of excitations related to multiple shape coexistence and excitation modes. Furthermore, the model self-consistently minimizes the 4D collective potential by including higher-order deformations. 

The theoretical framework is outlined in Sec. \ref{Sec:II}. Section \ref{Sec:III} presents illustrative calculations of the low-lying spectra, transition rates, and collective probability density distributions for $^{152}$Sm. Finally, a summary and outlook are provided in Section \ref{Sec:IV}.

\section{\label{Sec:II}Theoretical Framework}
\subsection{Triaxial quadrupole-octupole collective Hamiltonian}
The Triaxial Quadrupole-Octupole Collective Hamiltonian (TQOCH) provides a framework for describing nuclear excitations that involve quadrupole-octupole shape vibrations (both axial and triaxial) and collective rotations. The TQOCH is defined using the four deformation parameters $(\beta_{\lambda\mu})=(\beta_{20}, \beta_{22}, \beta_{30}, \beta_{32})$ and the three Euler angles $\Omega$ as its collective coordinates. Its general form reads:
\begin{align}
   \label{eq:TOCH}
{\hat H}_{\rm coll}(\beta_{\mu\nu}, \Omega) &= -\frac{\hbar^2}{2\sqrt{\omega{r}}}\sum\limits_{\lambda\mu, \lambda^
\prime\mu^\prime}\frac{\partial}{\partial \beta_{\lambda\mu}}\sqrt{\omega{r}}\left(B^{-1}\right)_{\lambda\mu, \lambda^
\prime\mu^\prime}\frac{\partial}{\partial \beta_{\lambda^
\prime\mu^\prime}}\nonumber\\
&\ \ \ \ +\frac{1}{2}\sum\limits_{k=1}^3\frac{\hat{J}_k^2}{{\cal I}_k}
+{V}_{\rm coll}(\mathbf{\beta}),
\end{align}
where the collective mass tensor is defined by
\begin{align}
B=\left(
\begin{array}{cccc}
B_{20,20} & B_{20,22} & B_{20,30} & B_{20,32}\\
B_{20,22} & B_{22,22} & B_{22,30} & B_{22,32}\\
B_{20,30} & B_{22,30} & B_{30,30} & B_{30,32}\\
B_{20,32} & B_{22,32} & B_{30,32} & B_{32,32}
\end{array}
\right).
\end{align}
$\omega={\rm det} B$, $r={\cal I}_1{\cal I}_2{\cal I}_3$, and ${\cal I}_k$ are the moments of inertia.
The corresponding volume element in the collective space reads
\begin{align}
\int{\rm d}\tau_{\rm coll}=\int\sqrt{\omega{r}}{\rm d}\beta_{20}{\rm d}\beta_{22}{\rm d}\beta_{30}{\rm d}\beta_{32}{\rm d}\Omega.
\end{align}

We solve the eigenvalue problem associated with the collective Hamiltonian in Eq.~(\ref{eq:TOCH}) by expanding its eigenfunctions in a complete basis. This basis, defined for each angular momentum $I$, is symmetrized with respect to $R_1$ and $R_2$ rotations~\cite{Kumar1NPA1967,RingBook} and the parity transformation, and is constructed as follows:
\begin{align}
\phi^\pi_n(\beta)|IMK\rangle
=&(\omega{r})^{-\frac{1}{4}} \prod\limits_{\lambda=2,3;\ \mu=0, 2}\phi_{n_{\lambda\mu}}(\beta_{\lambda\mu})\mid{IMK}\rangle,
\end{align}
where $\phi_{n_{\lambda\mu}}$ denotes the one-dimensional harmonic oscillator wave function for the variable $\beta_{\lambda\mu}$. The angular momentum component reads
\begin{align}
\mid{IMK}\rangle&=\sqrt{\frac{2I+1}{16\pi^2(1+\delta_{K0})}}\left[D^{I*}_{MK}(\Omega)+(-1)^{n_{30}+n_{32}+I}D^{I*}_{M-K}(\Omega)\right],
\end{align}
where $D^{I*}_{MK}$ is Wigner function and the projection $K$ is a non-negative even number. Furthermore, the quantum numbers $n_{22}$ and $n_{32}$ must satisfy the following condition:
\begin{align}
n_{22}+n_{32}=\left\{\begin{array}{ll}
\text{odd}&  \ \ \ \text{for}\ {K/2}=\text{odd}\\
\text{even}& \ \ \ \text{for}\ {K/2}=\text{even}
\end{array}\right..
\end{align}

Note that the symmetry under the $R_3$ rotation—a rotation of the x, y, z axes in the intrinsic frame-is broken in the present collective space because the octupole deformations are not fully included \cite{GreinerBook}. This breaking introduces spurious states, which can only be fully removed by restoring $R_3$ symmetry. In practice, however, we can effectively avoid low-lying spurious states by restricting the calculation to the subspace where $\beta_{20}\ge0$.

The diagonalization of the collective Hamiltonian Eq.~(\ref{eq:TOCH}) yields the energy spectrum $E^{I\pi}_\alpha$ and the corresponding eigenfunctions
\begin{align}\label{eq-Wav}
& \Psi^{IM\pi}_\alpha=\sum\limits_{K\in\text{even}}
\psi^{I\pi}_{\alpha{K}}(\beta_{20},\beta_{22},\beta_{30},\beta_{32})\mid{IMK}\rangle.
\end{align}
Using the collective wave functions~(\ref{eq-Wav}), various observables can be computed and compared with experimental values. For instance, the reduced electric multipole transition probability $B(E\lambda)$ reads
\begin{align}
B\left(E\lambda;\alpha{I}\rightarrow{\alpha^\prime}I^\prime\right)
=\frac{1}{2I+1}\left|\langle{\alpha^\prime}I^\prime\mid\mid\hat{\mathcal{M}}\left(E\lambda\right)
\mid\mid{\alpha}I\rangle\right|^2,
\end{align}
where $\hat{\mathcal{M}}(E\lambda)=\sum\limits_{\mu^\prime\in\text{even}}D^{\lambda^*}_{\mu\mu^\prime}Q^p_{\lambda\mu^\prime}$ is the electric multipole operator in the laboratory frame. $Q^p_{\lambda\mu^\prime}$ are the multipole moments for protons in the intrinsic frame~\cite{ZhouPS2016,LuPRC2014}. For electric dipole transitions, the intrinsic axially-symmetric dipole moment is defined \cite{ButlerRMP1996}:
\begin{align}
D_{10} &=\sqrt{\frac{3}{4\pi}}e\left(\frac{N}{A}\langle z_p\rangle-\frac{Z}{A}\langle z_n\rangle\right),
\end{align}
The electric monopole transition probability can be calculated from
\begin{align}
\rho^2(E0;\alpha{I}\rightarrow{\alpha^\prime I})
=\left|\frac{\langle\alpha{I}\mid\sum_ke_kr^2_k\mid\alpha^\prime I \rangle}{eR^2_0}\right|^2,
\end{align}
with $R_0=1.2\cdot A^{1/3}$ fm.
\subsection{Parameters of the collective Hamiltonian}

The dynamics of the collective Hamiltonian in Eq.~(\ref{eq:TOCH}) are governed by fourteen functions of the deformation parameters $(\beta_{\lambda\mu})$: the collective potential, ten mass parameters $B_{\lambda\mu,\lambda^\prime\mu^\prime}$, and three moments of inertia $\mathcal{I}_k\ (k=1,2,3)$. These functions are determined microscopically from multidimensionally constrained CDFT calculations. In the present study, the PC-PK1 energy density functional \cite{ZhaoPRC2010} is used for the particle-hole channel, while the Bardeen-Cooper-Schrieffer (BCS) approximation with a separable pairing force is employed for the particle-particle channel \cite{TianPLB2009,NIksicPRC2010}. The deformed single-particle equations are solved by expanding the wave functions in an axially deformed harmonic oscillator (ADHO) basis. For $^{152}$Sm, we use 12 major ADHO shells, which is sufficient for describing low-lying excitations. A detailed description of the multidimensionally constrained CDFT framework can be found in Ref.~\cite{LuPRC2014}.

The map of the collective energy surface as a function of $\beta_{\lambda\mu}$ is obtained by imposing constraints on the mass multipole moments $q_{\lambda\mu}$ \cite{RingBook}.
\begin{align}
& \langle E_{\rm CDFT}\rangle + \sum\limits_{\lambda\mu}C_{\lambda\mu}\left(\langle \hat{Q}_{\lambda\mu}\rangle - q_{\lambda\mu}\right)^2,
\label{constr}
\end{align}
where $E_{\rm CDFT}$ is the total energy, and  $\langle \hat{Q}_{\lambda\mu}\rangle$ denotes the expectation value of the mass quadrupole and octupole operators:
\begin{align}
\hat{Q}_{20}&=2z^2-x^2-y^2,\ \  \ \ \  \ \ \ \ \ \ \hat{Q}_{22}=x^2-y^2,\\
\hat{Q}_{30}&=z(2z^2-3x^2-3y^2),\ \ \ \hat{Q}_{32}=z(x^2-y^2) ~.
\end{align}
$q_{\lambda\mu}$ is the constrained value of the mutipole moment, and $C_{\lambda\mu}$ the corresponding stiffness constant \cite{RingBook}.

The single-nucleon wave functions, energies and occupation probabilities, generated from constrained CDFT calculations, provide the microscopic input for the parameters of the collective Hamiltonian. The moments of inertia are computed using the Inglis-Beliaev formula \cite{InglisPR1956,BeliaevNP1961}
\begin{align}
\label{eq:MOI}
\mathcal{I}_k &= \sum_{i,j}\frac{| \langle i |\hat{J}_k | j  \rangle |^2}{E_i+E_j}\left(u_iv_j-v_iu_j \right)^2,
\end{align}
where $k$ denotes the axis of rotation, and the summation runs over the proton and neutron quasiparticle states. The mass parameters are calculated in the cranking approximation \cite{GirodNPA1979}:
\begin{align}
B_{\lambda\mu,\lambda^\prime\mu^\prime}=\hbar^2\left[\mathcal{M}^{-1}_{(1)}\mathcal{M}_{(3)}\mathcal{M}^{-1}_{(1)}\right]_{\lambda\mu,\lambda^\prime\mu^\prime},
\end{align}
with
\begin{align}
\mathcal{M}_{(n){\lambda\mu,\lambda^\prime\mu^\prime}}&=\sum\limits_{ij}\frac{\left\langle i\right|\hat{Q}_{\lambda\mu}\left| j\right\rangle\left\langle j\right|\hat{Q}_{\lambda^\prime\mu^\prime}\left| i\right\rangle}{(E_i+E_j)^n}\left(u_i v_j+ v_i u_j \right)^2.
\end{align}
The collective potential $V_{\rm coll}$ in Eq.~(\ref{eq:TOCH}) is obtained by subtracting the vibrational and rotational zero-point energy (ZPE) corrections from the total mean-field energy:
\begin{align}
{V}_{\rm coll} =  \langle E_{\rm CDFT}\rangle - \Delta V_{\rm vib} - \Delta V_{\rm rot}.
\end{align}
The vibrational and rotation ZPE corrections are computed in the cranking approximation \cite{GirodNPA1979}
\begin{align}
\Delta{V}_{\rm vib}=\frac{1}{4}\text{Tr}\left(\mathcal{M}^{-1}_{(3)}\mathcal{M}_{(2)}\right),
\end{align}
and
\begin{align}
\Delta{V}_{\rm rot} = \sum\limits_{k=1}^3\frac{\langle\hat{J}_k^2\rangle}{\mathcal{I}_k}.
\end{align}

\section{\label{Sec:III}Results and discussion}

\begin{figure*}[ht]
\centering{\includegraphics[width=0.7\textwidth]{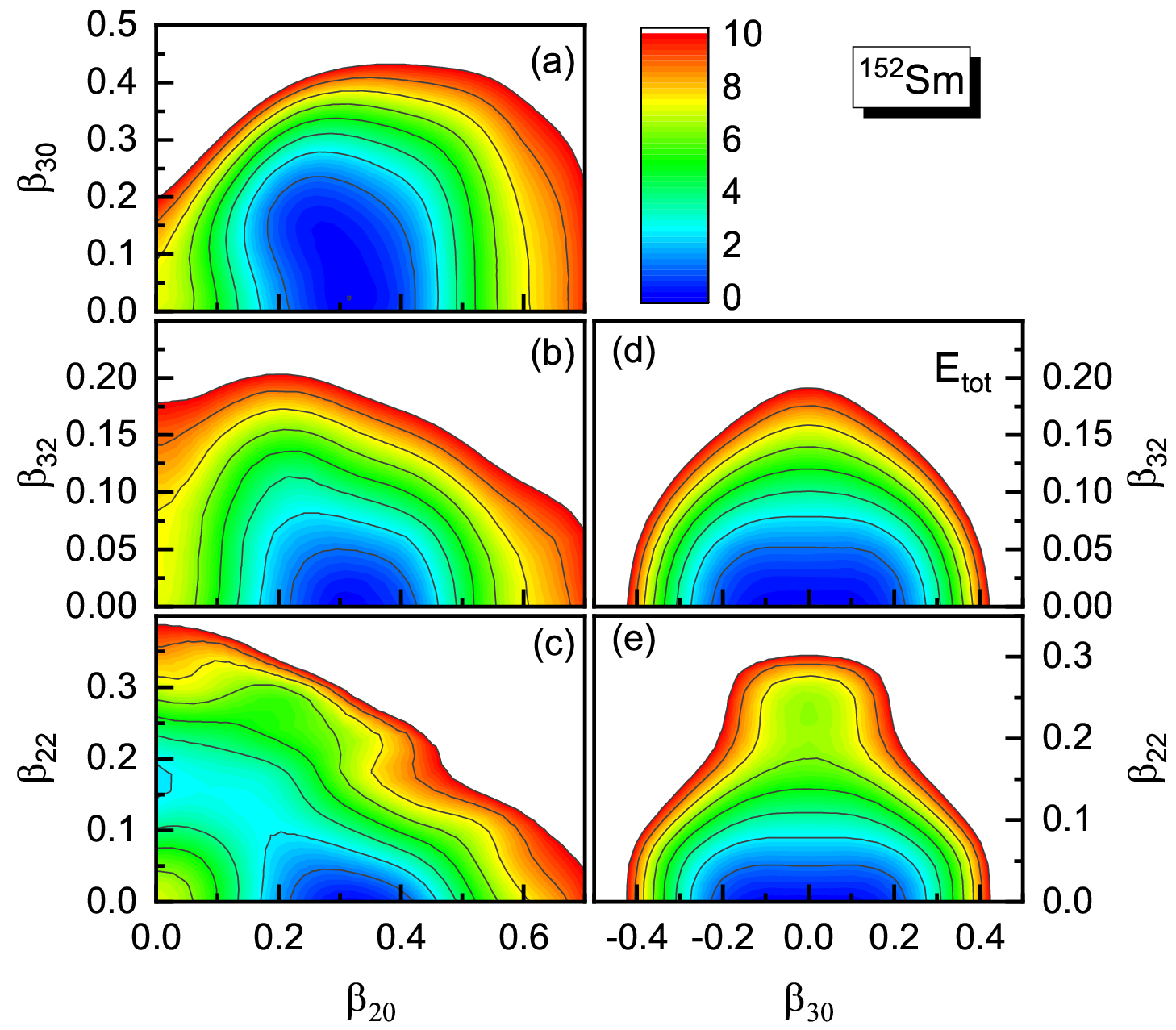}}
  \caption{\label{PES-Sm152}(Color online) Deformation energy surface (DES) of $^{152}$Sm computed using the multidimensionally constrained CDFT based on the PC-PK1 functional and a separable pairing interaction (a-e). Two-dimensional projections of the DES near the global minimum on the ($\beta_{20},\beta_{30}$), ($\beta_{20},\beta_{32}$), ($\beta_{20},\beta_{22}$),  ($\beta_{30},\beta_{22}$), and ($\beta_{30},\beta_{32}$) planes. The contours join points on the surface with the same energy.}
\end{figure*}

\begin{figure*}[ht]
 \includegraphics[width=0.92\textwidth]{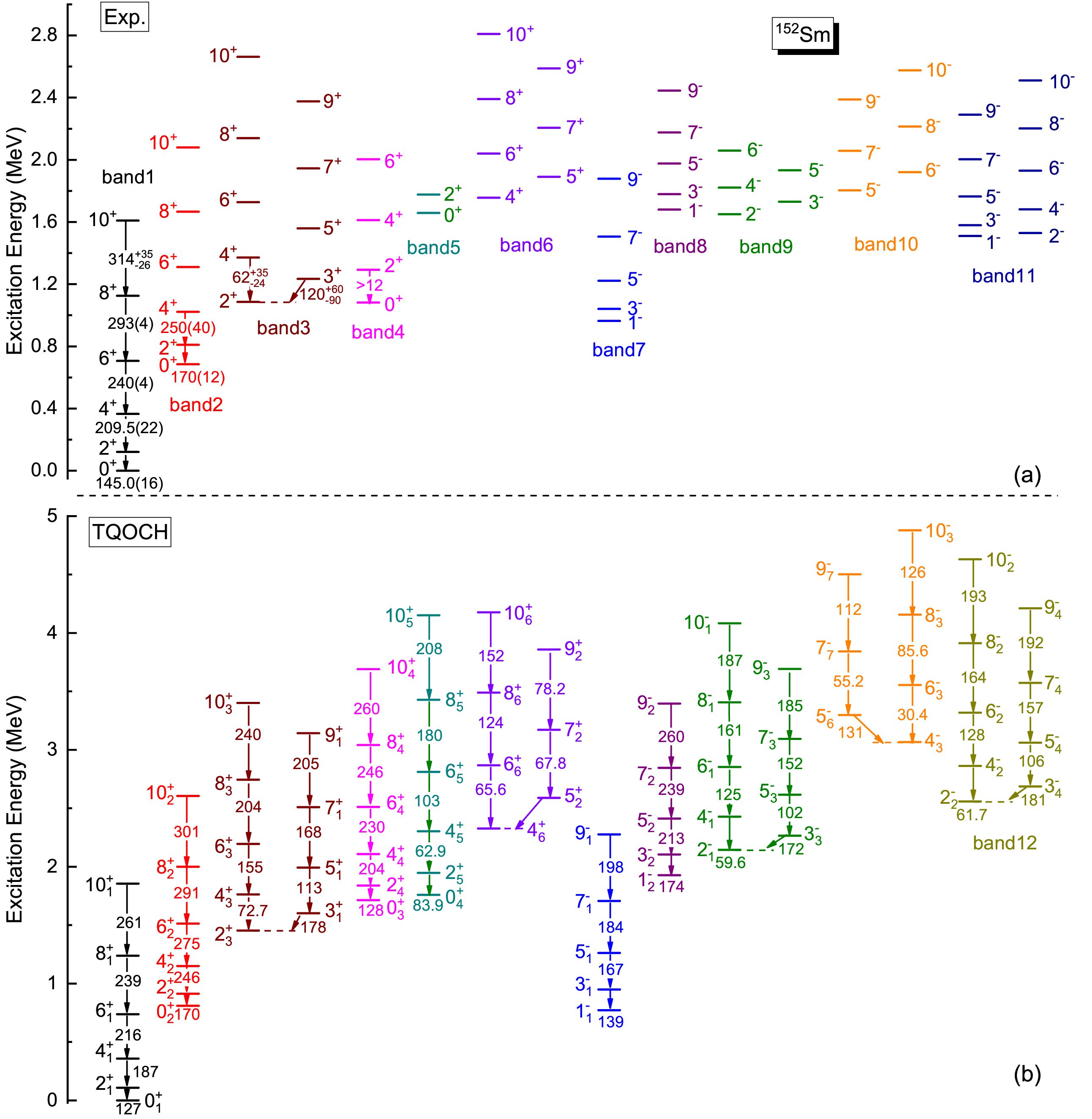}
  \caption{\label{Spectra-Sm152}(Color online) Low-energy positive- and negative-parity bands, and the corresponding $B(E2)$ values (in Weisskopf units) for intraband transitions in $^{152}$Sm. The TQOCH results (panel b) are shown in comparison with  available data (panel a)~\cite{NNDC,GarrettAIPCP2009}.}
\end{figure*}

To demonstrate the capabilities of our new implementation, we apply it to calculate the deformation energy surface (DES) and collective excitation spectra of $^{152}$Sm. This nucleus is an ideal test case due to the abundance of experimental data available for its low-lying positive- and negative-parity excitation spectra. Figure \ref{PES-Sm152} presents the DES of $^{152}$Sm, projected onto the ($\beta_{20},\beta_{30}$), ($\beta_{20},\beta_{32}$),
($\beta_{20},\beta_{22}$),   ($\beta_{30},\beta_{22}$), and ($\beta_{30},\beta_{32}$) planes, intersecting at the global minimum of $(\beta_{20},\beta_{22},\beta_{30},\beta_{32})\approx(0.30,0.00,0.00,0.00)$. Notably, panel (a) reveals that the DES is very soft along both the axially symmetric quadrupole $\beta_{20}$ and octupole $\beta_{30}$ deformation coordinates. The DES remains particularly flat with respect to octupole deformation up to $\beta_{30}\approx 0.2$, which aligns with previous axially symmetric DFT calculations using both relativistic ~\cite{ZhangPRC2010,NomuraPRC2014,XiaPRC2017} and non-relativistic~\cite{NomuraPRC2015} functionals. In contrast, panels (b-e) show that the DES is stiff with respect to nonaxial deformations. The profile in the ($\beta_{20},\beta_{22}$) plane is consistent with our earlier findings. Most significantly, this work provides the first microscopic description of the DES along the $\beta_{32}$ coordinate for $^{152}$Sm
 and the corresponding vibrational excitations (see Figs. \ref{Spectra-Sm152} and \ref{Wav-1-1}).

\begin{figure}[ht]
 \includegraphics[width=0.45\textwidth]{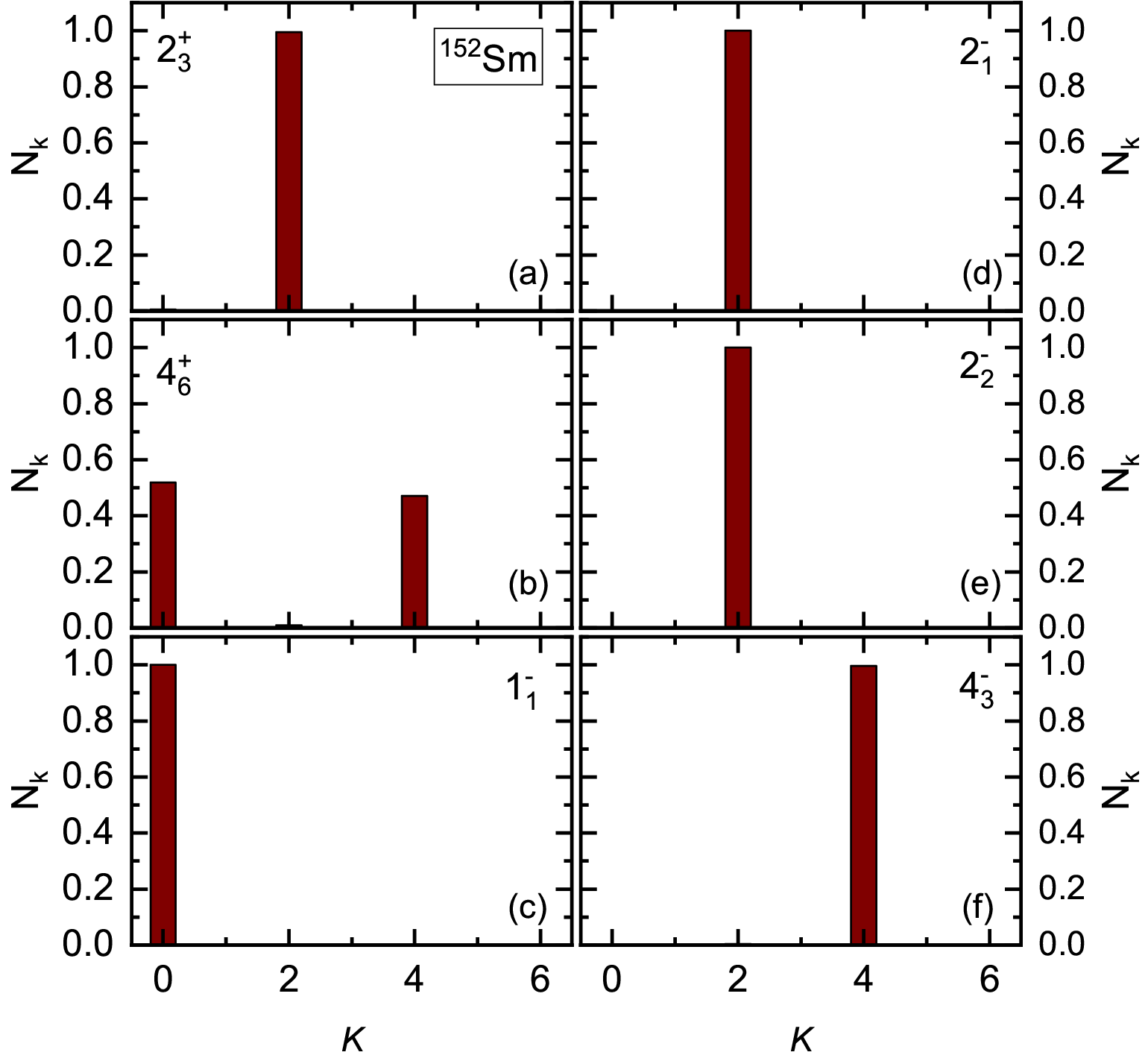}
  \caption{\label{Nk-Sm152}(Color online) Distribution of $K$-components (projection of the angular momentum on the intrinsic z-axis) in the collective wave functions of the states $2^+_3$, $4^+_6$, $1^-_1$, $2^-_{1}$, $2^-_{2}$, and $4^-_3$.}
\end{figure}

The diagonalization of the resulting Hamiltonian yields the excitation energies and collective wave functions for each total angular momentum and parity, $I^\pi$. As an illustrative example, Fig.~\ref{Spectra-Sm152} presents the first six positive-parity and five negative-parity low-lying bands in $^{152}$Sm, along with their calculated intraband $B(E2)$ values, compared to available experimental data~\cite{NNDC}. The corresponding interband transitions-$\rho^2(E0)$, $B(E1)$, $B(E2)$, and $B(E3)$-are quantified in Tables \ref{Tab-E2transition} and \ref{Tab-E1transition}. Furthermore, Fig.~\ref{Nk-Sm152} displays the distribution of K components (the projection of the angular momentum on the intrinsic z-axis) for the bandheads $2^+_3$, $4^+_6$, $1^-_1$, $2^-_{1,2}$, and $4^-_3$. 

\begin{figure*}[ht]
  \centering{\includegraphics[width=0.46\textwidth]{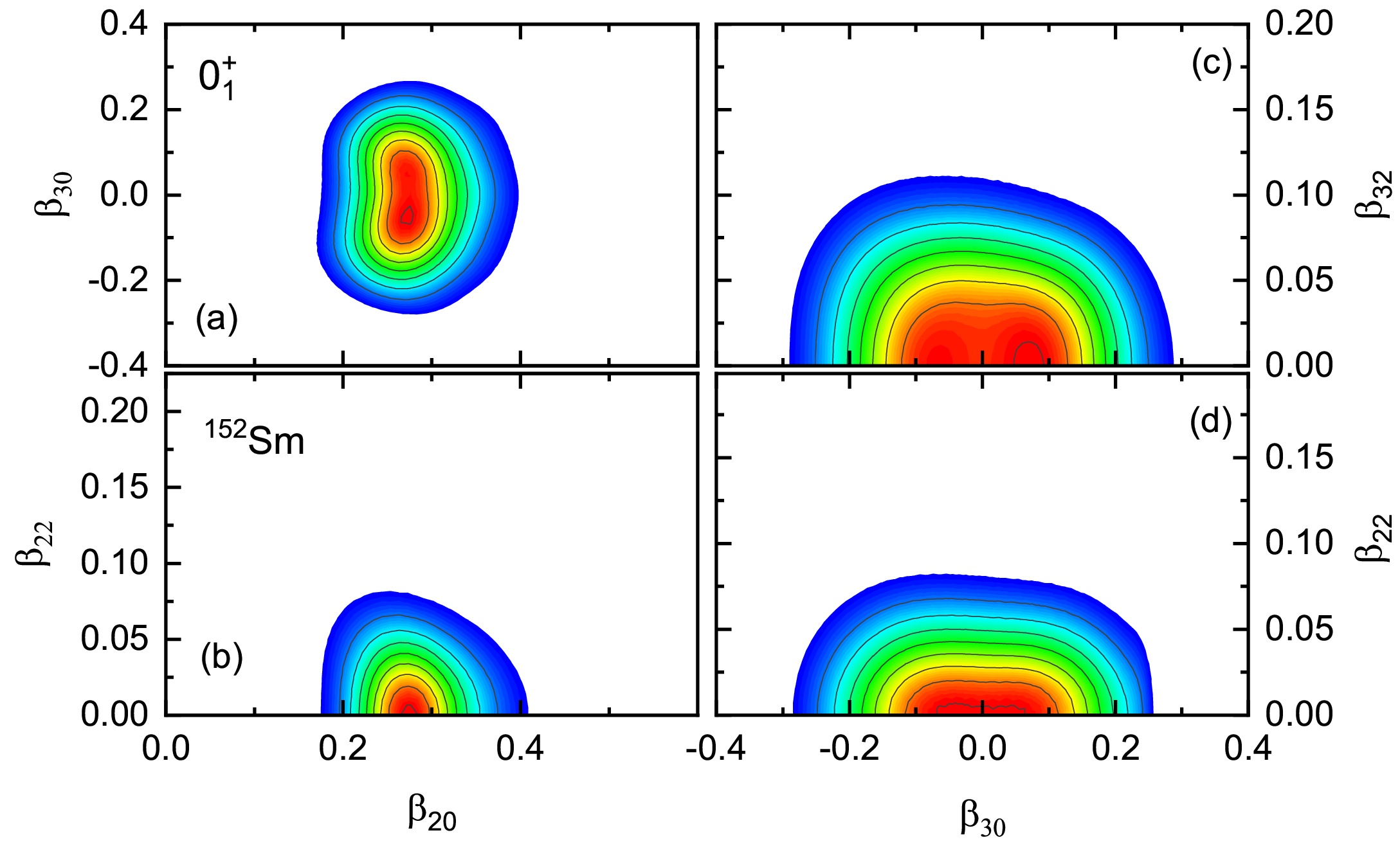}}
  \includegraphics[width=0.46\textwidth]{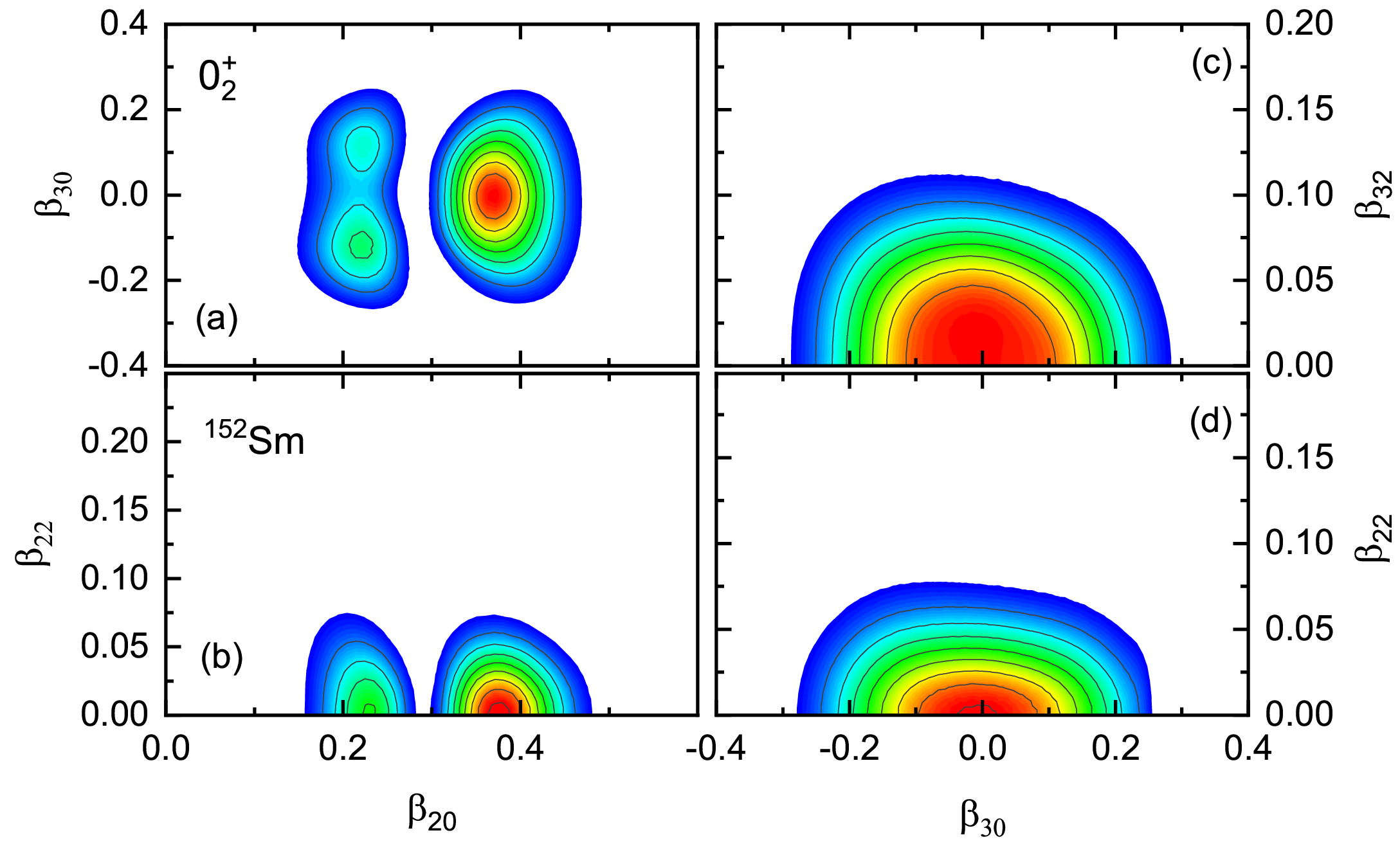}
  \includegraphics[width=0.46\textwidth]{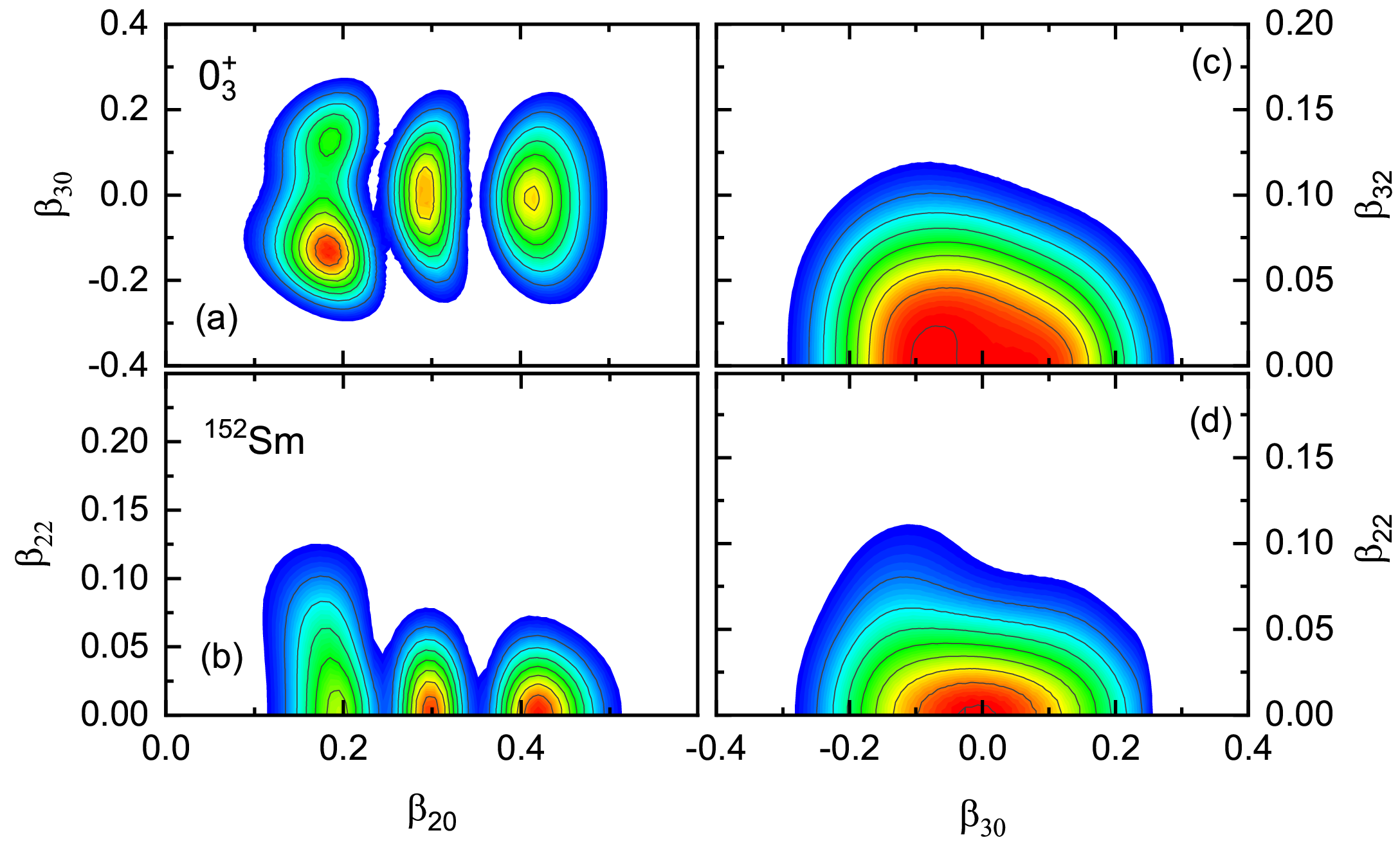}
  \includegraphics[width=0.46\textwidth]{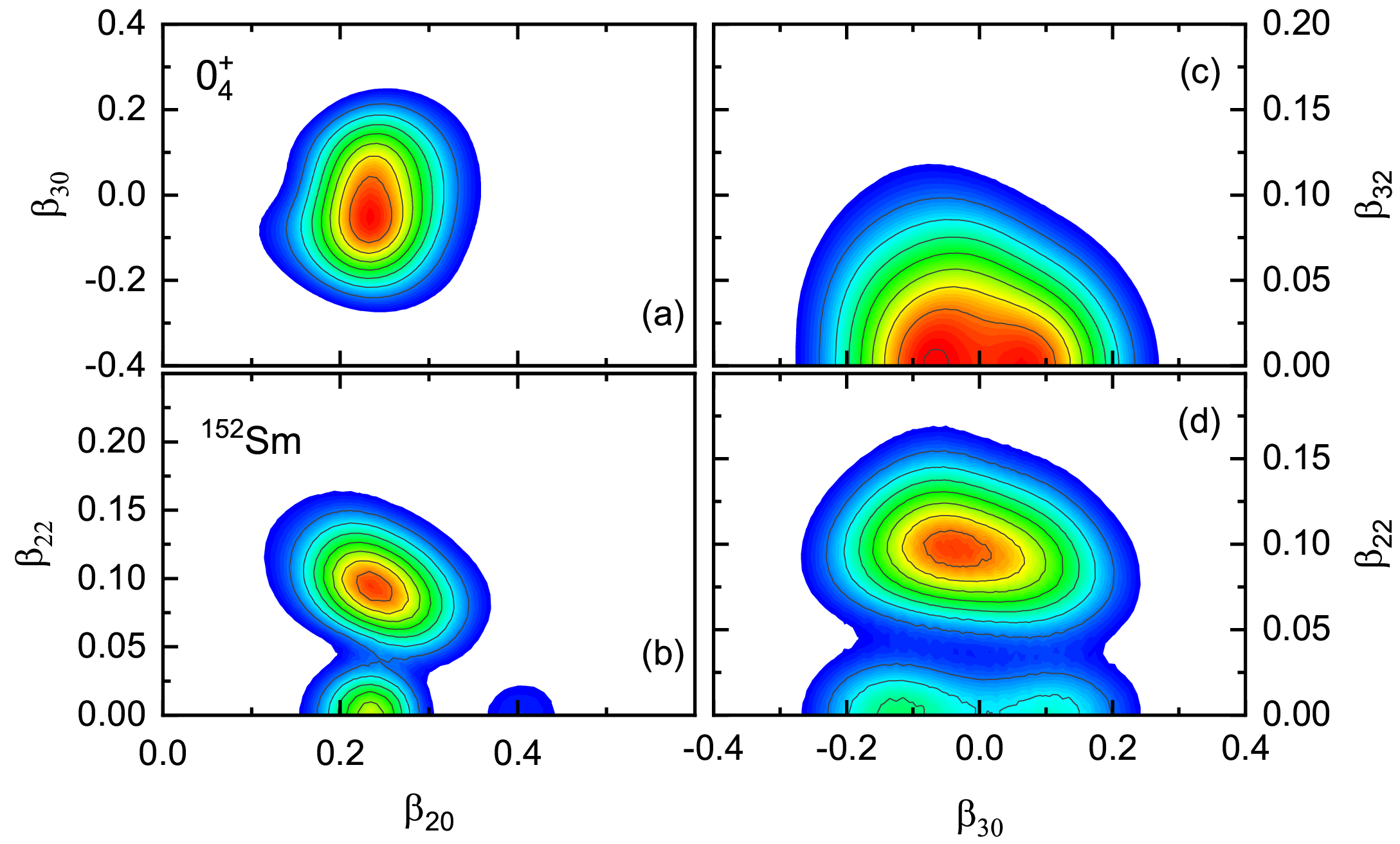}
  \includegraphics[width=0.46\textwidth]{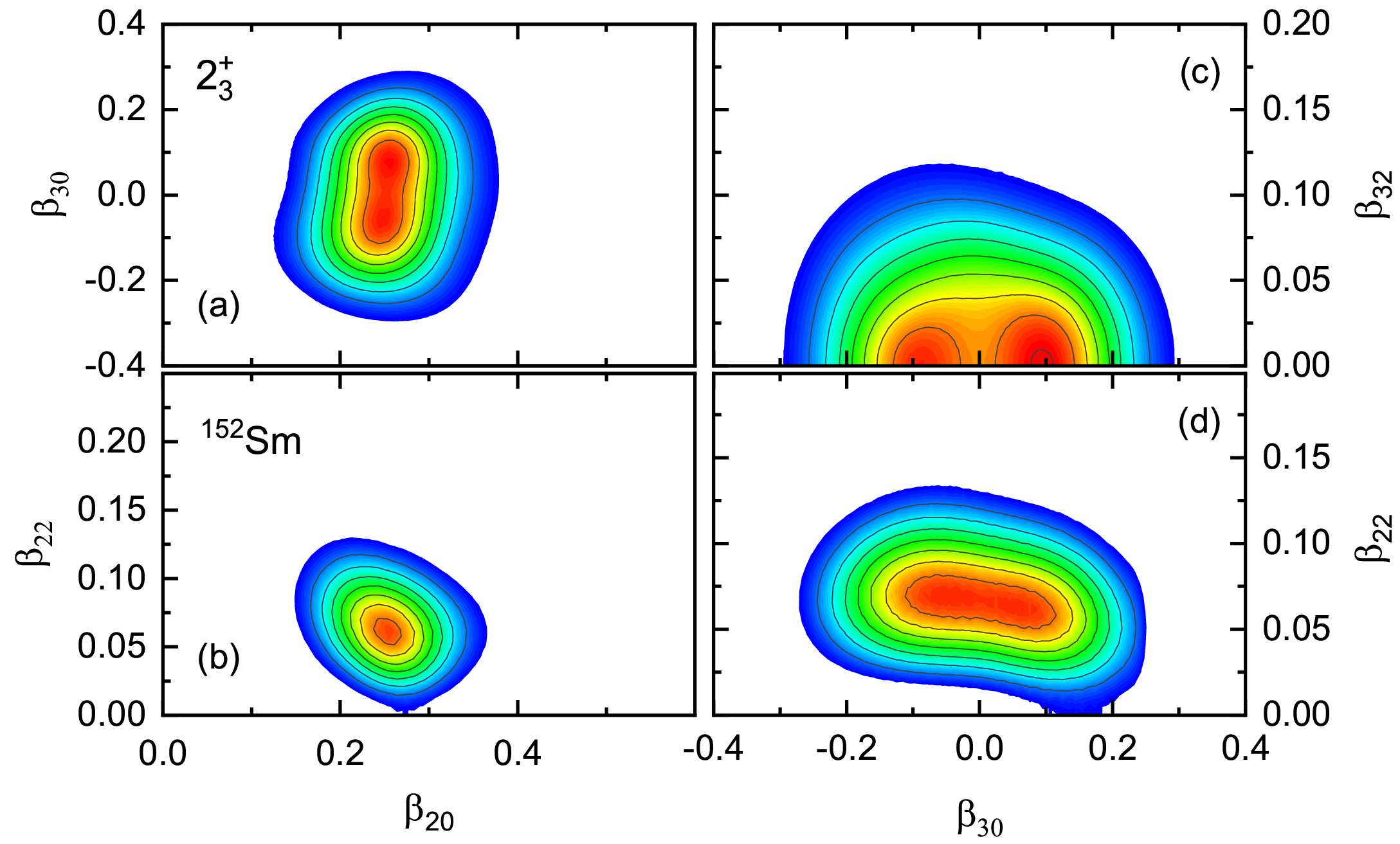}
  \includegraphics[width=0.46\textwidth]{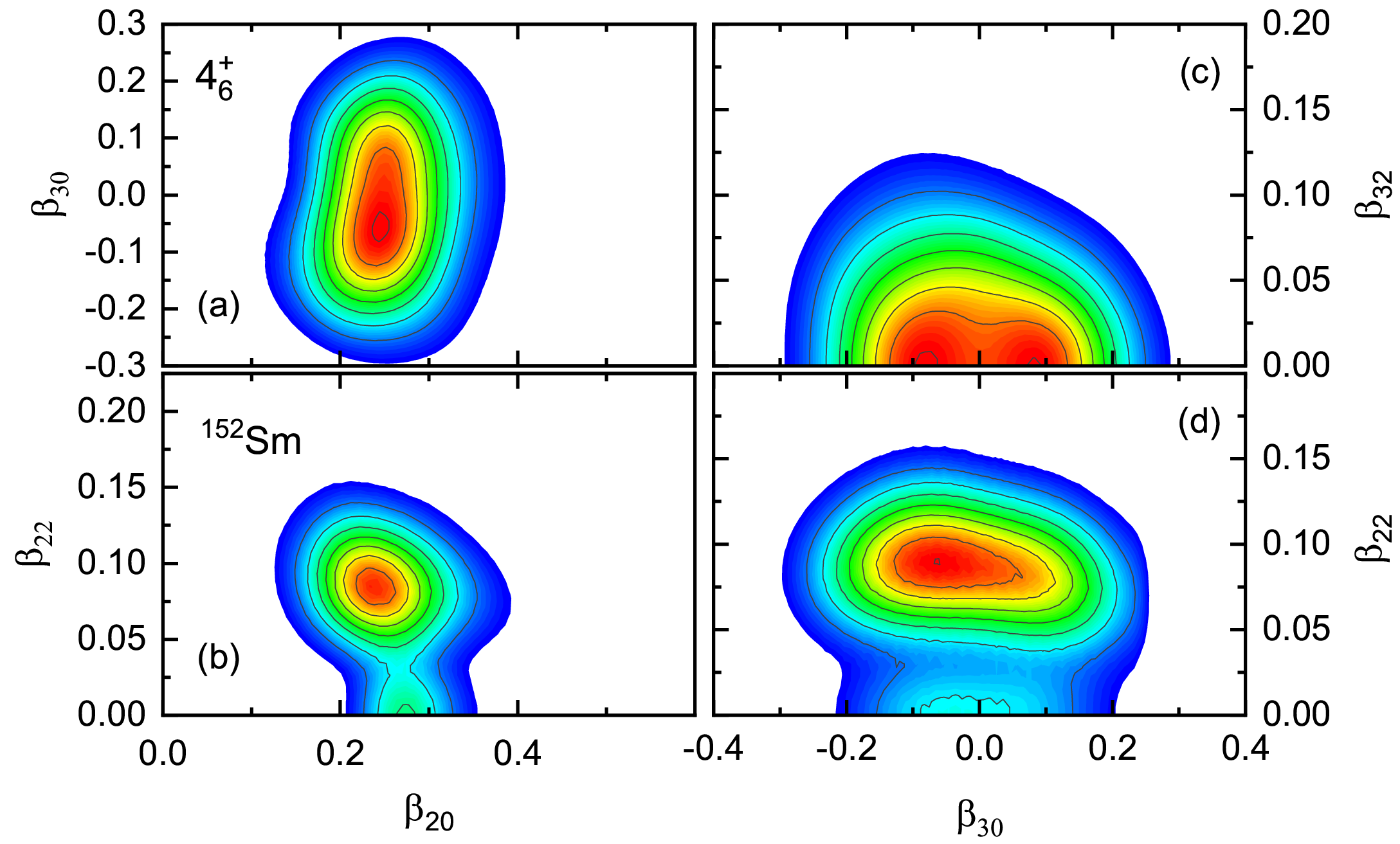}
  \caption{\label{Wav-0+1}(Color online) The probability density distributions for the states $0^+_{1,2,3,4}$, $2^+_3$, and $4^+_6$ of $^{152}$Sm in the ($\beta_{20},\beta_{22}$), $(\beta_{20},\beta_{30})$, ($\beta_{30},\beta_{22}$), $(\beta_{30},\beta_{32})$ planes.}
\end{figure*}

\begin{figure*}[ht]
  \centering{\includegraphics[width=0.46\textwidth]{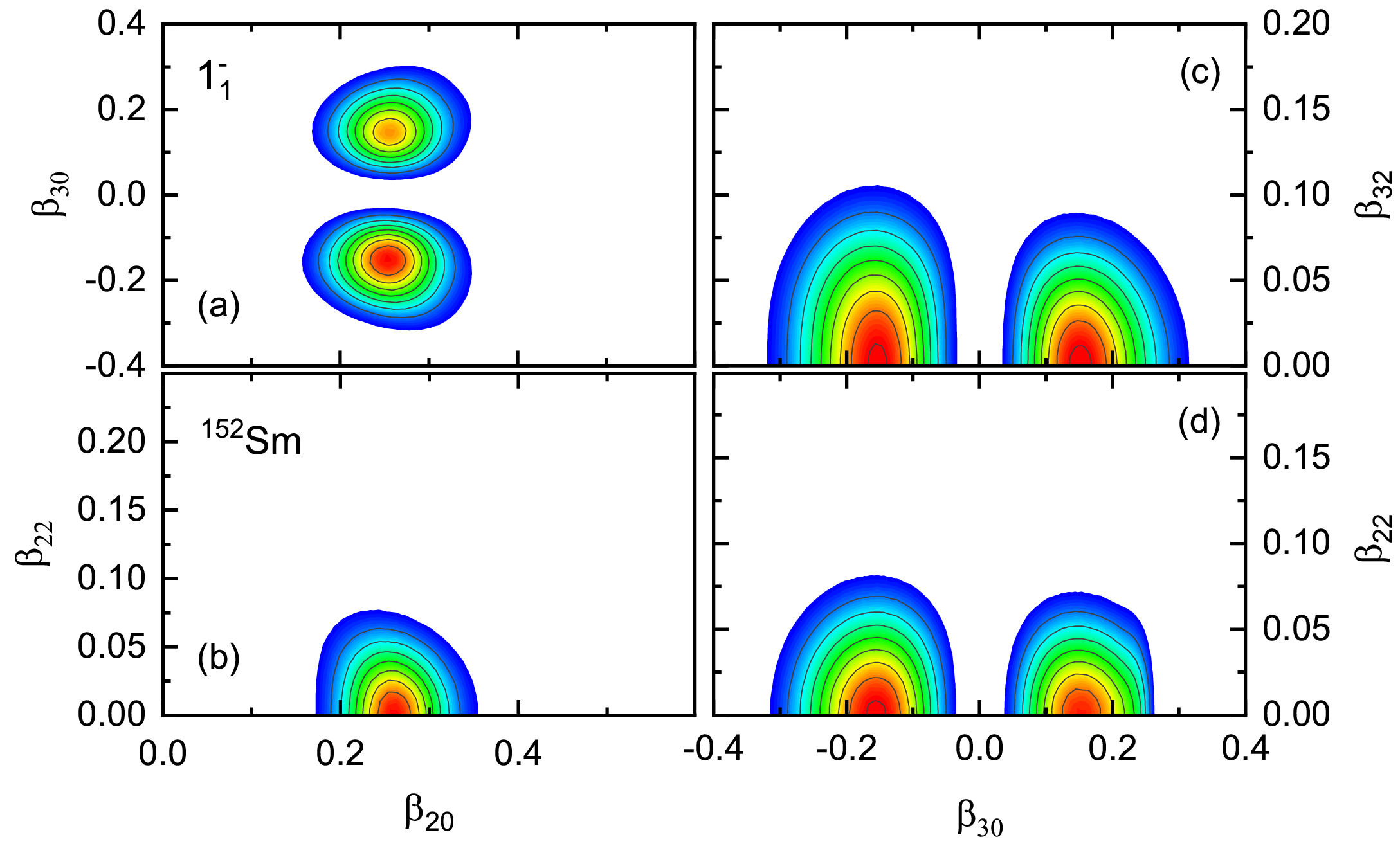}}
  \includegraphics[width=0.46\textwidth]{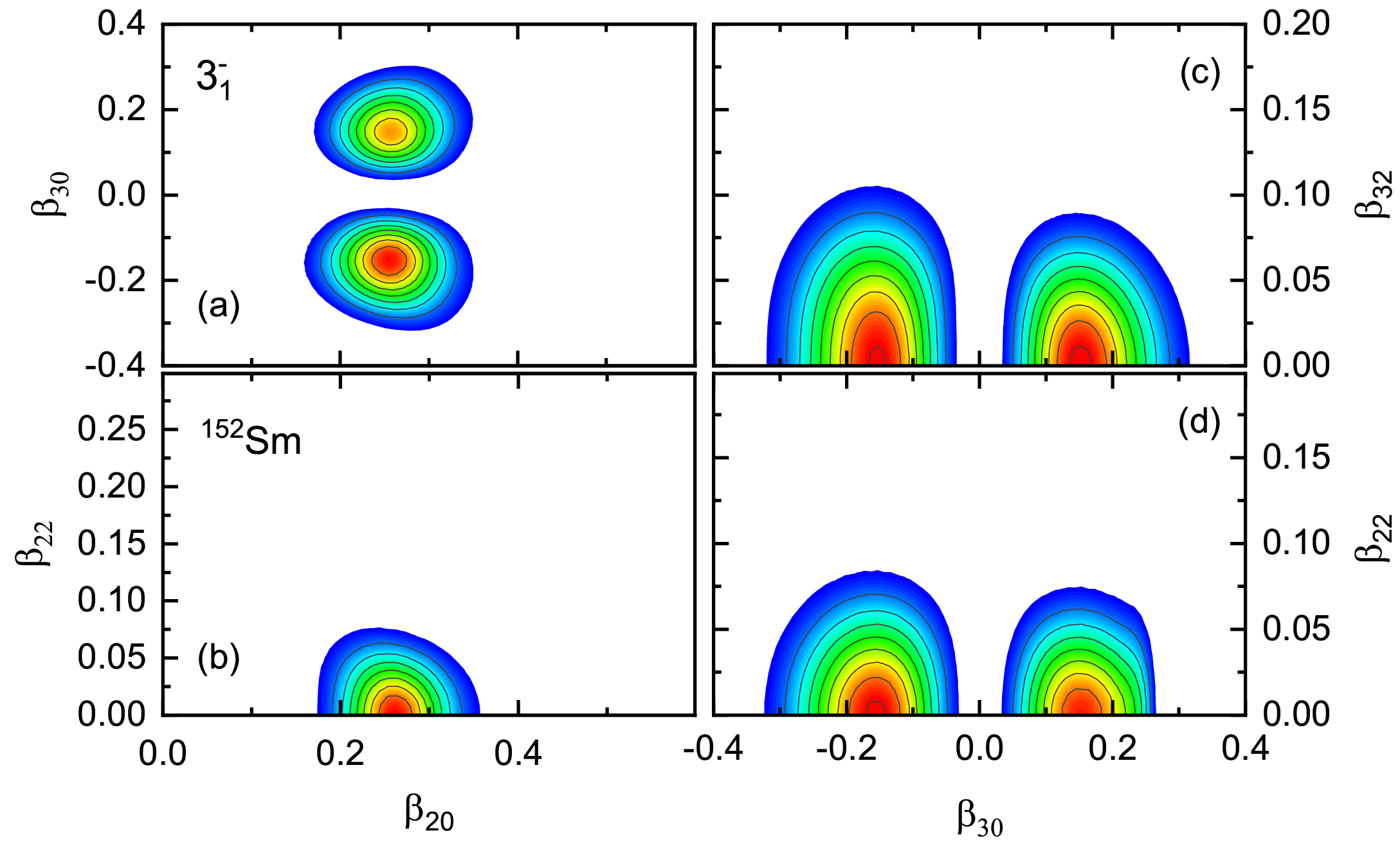}
  \includegraphics[width=0.46\textwidth]{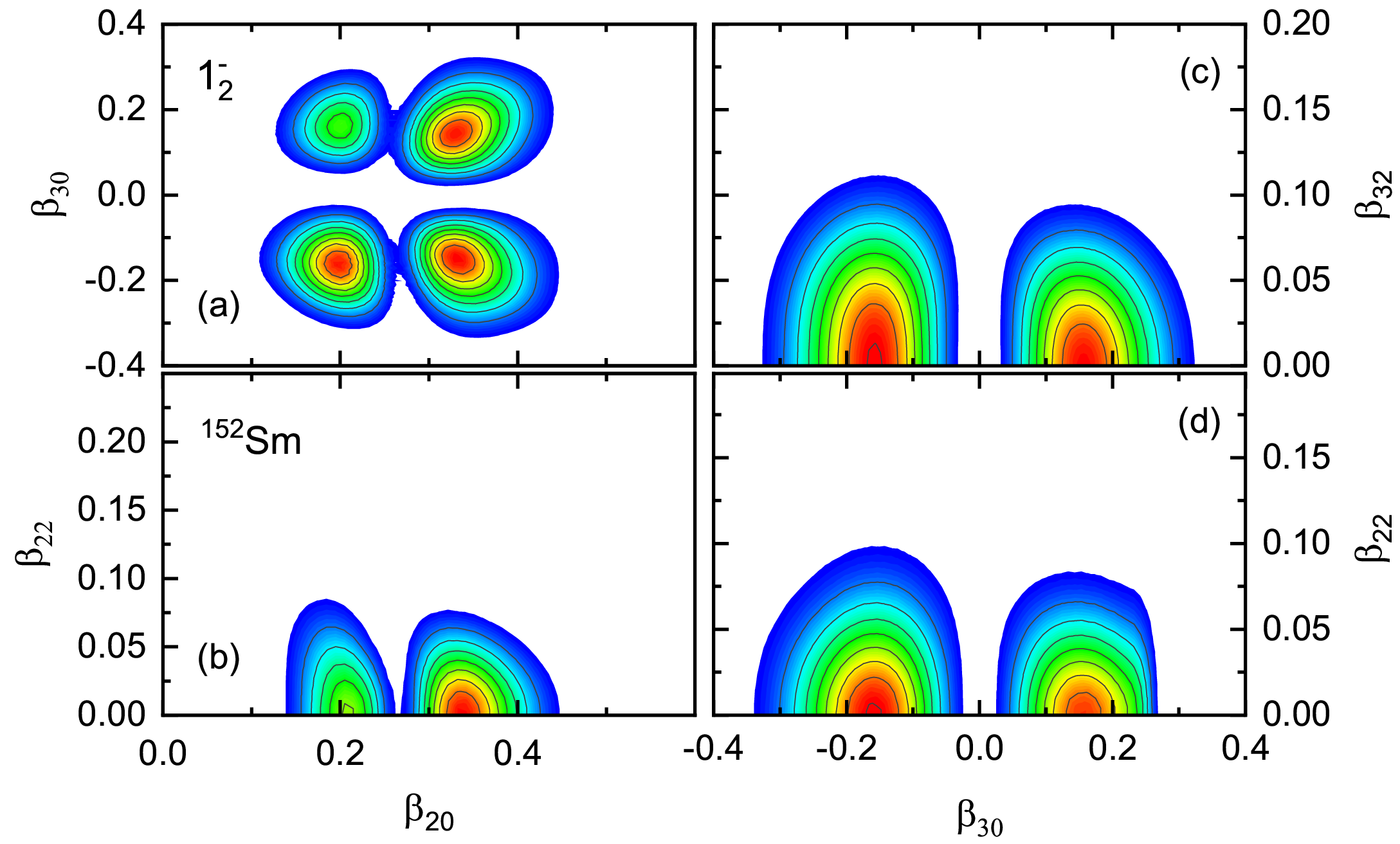}
  \includegraphics[width=0.46\textwidth]{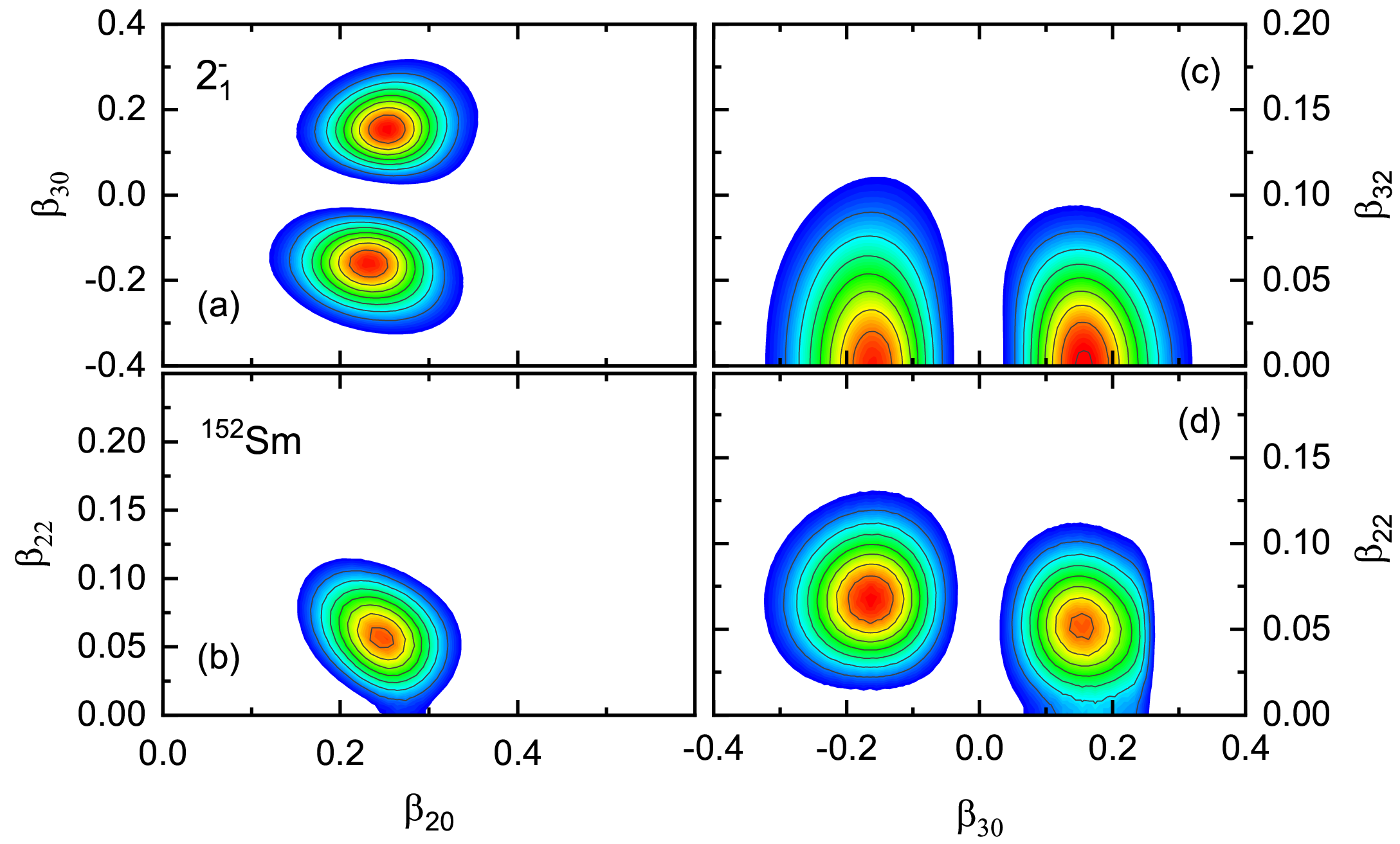}
  \includegraphics[width=0.46\textwidth]{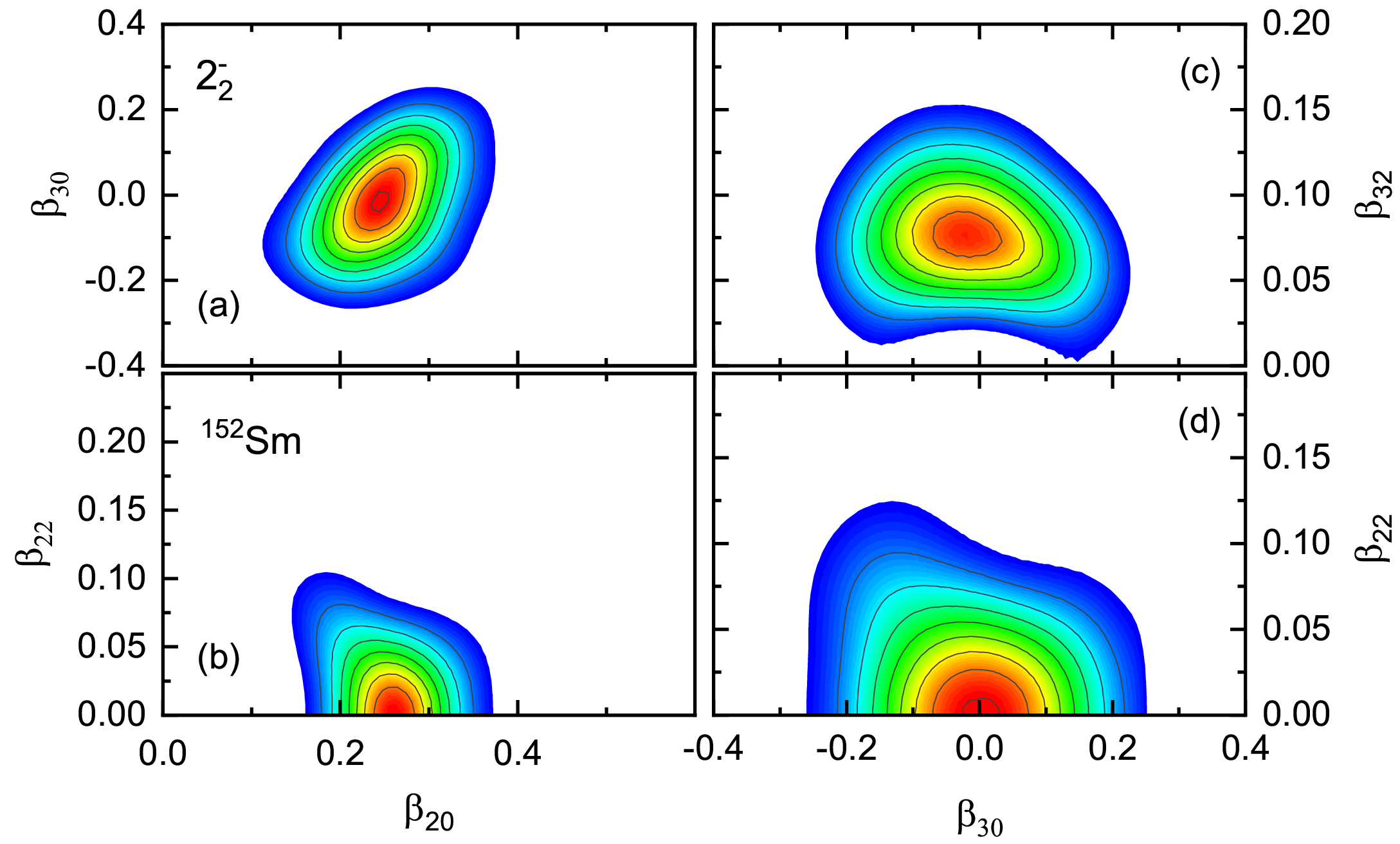}
  \includegraphics[width=0.46\textwidth]{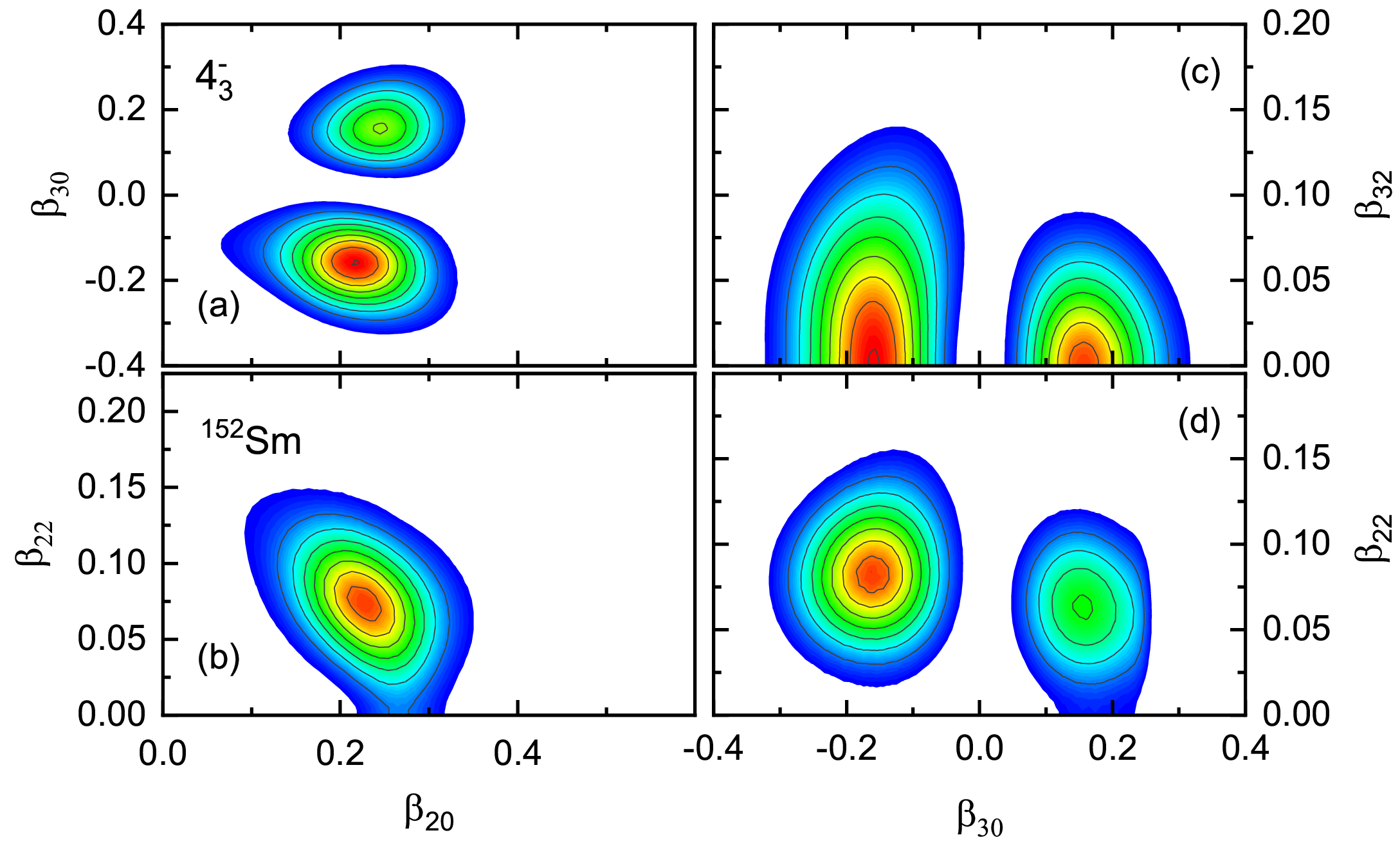}
  \caption{\label{Wav-1-1}(Color online) Same as in the caption to Fig. \ref{Wav-0+1} but for the states $1^-_{1,2}$, $2^-_{1,2}$, $3^-_1$, and $4^-_3$ of $^{152}$Sm.}
\end{figure*}

Our microscopic TQOCH calculations successfully reproduce the structure of both the positive- and negative-parity bands. We note, however, that the theoretical bands are generally overextended in energy, a result of the underestimation of moments of inertia inherent to the cranking approximation \cite{RingBook,NiksicPRC2019}. The theoretical results for the positive-parity ground-state band, the $0^+_2$ band (band 2), and the $\gamma$-band (band 3) show good agreement with available experimental data. Notably, the new model provides a significantly improved description of the $B(E2)$ values for interband transitions between these bands (see Table \ref{Tab-E2transition}) compared to our previous triaxial quadrupole collective Hamiltonian \cite{LiPRC2009}. This improvement also extends to the $0^+_3$ band, for which the model successfully reproduces the large $B(E2; 0^+_3\to 2^+_2)$ value, despite predicting the overall energy of the band too high. However, the $\rho^2(E0)$ value between the $0^+_3$ band (band 4) and the $0^+_2$ band (band 2) is overestimated by an order of magnitude. We anticipate that these discrepancies could be mitigated by including the degree of pairing vibration, as demonstrated in our previous work on the quadrupole collective Hamiltonian \cite{XiangPRC2020,XiangPRC2024}. The theoretical $0^+_4$ band is consistent with its experimental counterpart. However, due to limited transition data, we will analyze its structure based on the calculated probability density distribution in Fig.~\ref{Wav-0+1}. Finally, band 6 (band-head $4^+_6$) exhibits a structure and intraband transitions similar to those of the $\gamma$-band (band 3). Its wave functions are dominated by $K = 0$ and $K = 4$ components, identifying it as a second $\gamma$-band.

\begin{table}[htbp]
\tabcolsep=5pt   
\begin{center}
\caption{\label{Tab-E2transition}  The $10^3\rho^2(E0)$ and $B(E2)$ (in Weisskopf units) values for interband transitions in  $^{152}$Sm. The TQOCH results are compared to available data~\cite{NNDC,KibediPPNP2022}.}
\begin{tabular}{ccc}
\hline\hline
$B(E\lambda;I^\pi_\alpha\rightarrow I^{\prime\pi}_{\alpha^\prime})$          & Exp. (W.U.)   &  TQOCH (W.U.)\\
\hline
$10^3\rho^2(E0; 0^+_{\text{band2}}\rightarrow0^+_{\text{band1}})$             & 53(6)         &   101     \\
$10^3\rho^2(E0; 2^+_{\text{band2}}\rightarrow2^+_{\text{band1}})$             & 76(6)         &   104     \\
$10^3\rho^2(E0; 4^+_{\text{band2}}\rightarrow4^+_{\text{band1}})$             & 90(15)        &   109     \\
$10^3\rho^2(E0; 0^+_{\text{band4}}\rightarrow0^+_{\text{band1}})$             & 0.9(5)        &   0.06   \\
$10^3\rho^2(E0; 0^+_{\text{band4}}\rightarrow0^+_{\text{band2}})$             & 17(+8-7)      &   214     \\
$10^3\rho^2(E0; 2^+_{\text{band4}}\rightarrow2^+_{\text{band2}})$             & 11(+7-6)      &   221     \\
$B(E2; 0^+_{\text{band2}}\rightarrow2^+_{\text{band1}})$     &  33.3(12)    &  26.6 \\
$B(E2; 2^+_{\text{band2}}\rightarrow4^+_{\text{band1}})$     &  18.0(12)    &  17.3 \\
$B(E2; 2^+_{\text{band2}}\rightarrow2^+_{\text{band1}})$     &  5.7(4)      &  6.24 \\
$B(E2; 2^+_{\text{band2}}\rightarrow0^+_{\text{band1}})$     &  0.94(6)     &  3.25 \\
$B(E2; 4^+_{\text{band2}}\rightarrow6^+_{\text{band1}})$     &  17(3)       &  18.4 \\
$B(E2; 4^+_{\text{band2}}\rightarrow4^+_{\text{band1}})$     &  5.0(+10-7)  &  5.96 \\
$B(E2; 4^+_{\text{band2}}\rightarrow2^+_{\text{band1}})$     &  0.74(12)    &  3.26 \\
$B(E2; 2^+_{\text{band3}}\rightarrow0^+_{\text{band2}})$     &  0.026(4)    &  0.10 \\
$B(E2; 2^+_{\text{band3}}\rightarrow4^+_{\text{band1}})$     &  0.56(8)     &  0.39 \\
$B(E2; 2^+_{\text{band3}}\rightarrow2^+_{\text{band1}})$     &  7.4(1)      &  6.25 \\
$B(E2; 2^+_{\text{band3}}\rightarrow0^+_{\text{band1}})$     &  2.9(4)      &  3.61 \\
$B(E2; 3^+_{\text{band3}}\rightarrow4^+_{\text{band1}})$     &  7.4(+16-11) &  3.66 \\
$B(E2; 3^+_{\text{band3}}\rightarrow2^+_{\text{band1}})$     &  6.8(+15-11) &  6.43 \\
$B(E2; 4^+_{\text{band3}}\rightarrow2^+_{\text{band2}})$     &  0.30(+18-13)&  1.9E-5\\
$B(E2; 4^+_{\text{band3}}\rightarrow6^+_{\text{band1}})$     &  0.9(+6-4)   &  0.81 \\
$B(E2; 4^+_{\text{band3}}\rightarrow4^+_{\text{band1}})$     &  7(+4-3)     &  7.27 \\
$B(E2; 4^+_{\text{band3}}\rightarrow2^+_{\text{band1}})$     &  0.7(+5-3)   &  1.60 \\
$B(E2; 0^+_{\text{band4}}\rightarrow2^+_{\text{band2}})$     &  34(+23-11)  &  48.1 \\
$B(E2; 0^+_{\text{band4}}\rightarrow2^+_{\text{band1}})$     &  0.80(+53-23)&  1.44 \\
$B(E2; 2^+_{\text{band4}}\rightarrow4^+_{\text{band2}})$     &  $>6.2$      &  28.8 \\
$B(E2; 2^+_{\text{band4}}\rightarrow4^+_{\text{band1}})$     &  $>0.46$     &  0.54 \\
$B(E2; 3^-_{\text{band9}}\rightarrow1^-_{\text{band7}})$     &  6.9(+12-11) &  2.50\\
\hline\hline
\end{tabular}
\end{center}
\end{table}

\begin{table}[htbp]
\tabcolsep=5pt   
\begin{center}
\caption{\label{Tab-E1transition}  The  reduced electric $E1$ and $E3$ transition strengths $B(E1)$ and $B(E3)$ (in Weisskopf units) for transitions in $^{152}$Sm. The TQOCH results are shown in comparison with available data~\cite{NNDC,KibediADNDT2002}.}
\begin{tabular}{ccc}
\hline\hline
$B(E\lambda;I^\pi_\alpha\rightarrow I^{\prime\pi^\prime}_{\alpha^\prime})$ & Exp. (W.U.)  &  TQOCH (W.U.)\\
\hline
$B(E1; 1^-_{\text{band7}}\rightarrow0^+_{\text{band1}})$     &  0.0058(5)      &  0.0074  \\
$B(E1; 1^-_{\text{band7}}\rightarrow2^+_{\text{band1}})$     &  0.0106(9)      &  0.0150  \\
$B(E1; 3^-_{\text{band7}}\rightarrow2^+_{\text{band1}})$     &  0.0081(15)     &  0.0100  \\
$B(E1; 3^-_{\text{band7}}\rightarrow4^+_{\text{band1}})$     &  0.0082(16)     &  0.0138  \\
$B(E1; 5^-
_{\text{band7}}\rightarrow4^+_{\text{band1}})$     &  0.0042(+9-8)   &  0.0116\\
$B(E1; 5^-_{\text{band7}}\rightarrow6^+_{\text{band1}})$     &  0.0043(+9-8)   &  0.0145  \\
$B(E1; 1^-_{\text{band7}}\rightarrow2^+_{\text{band2}})$     &  0.000225(27)   &  6.3E-5  \\
$B(E1; 4^+_{\text{band3}}\rightarrow3^-_{\text{band7}})$     &  0.000064(+37-25)&0.0000312  \\
$B(E1; 4^+_{\text{band3}}\rightarrow5^-_{\text{band7}})$     &  0.000066(+38-26)&0.0000465   \\
$B(E1; 0^+_{\text{band4}}\rightarrow1^-_{\text{band7}})$     &  0.00063(+43-19)&  0.00035 \\
$B(E1; 1^-_{\text{band8}}\rightarrow0^+_{\text{band1}})$    &  4.4E-5(4)      & 0.00568 \\
$B(E1; 1^-_{\text{band8}}\rightarrow2^+_{\text{band1}})$    &  7.7E-5(+7-6)   & 0.0118 \\
$B(E1; 1^-_{\text{band8}}\rightarrow0^+_{\text{band2}})$    &  0.00240(+22-19)& 0.00461 \\
$B(E1; 1^-_{\text{band8}}\rightarrow2^+_{\text{band2}})$    &  0.0049(4)      & 0.00879 \\
$B(E1; 3^-_{\text{band8}}\rightarrow2^+_{\text{band2}})$    &  0.0028(+6-5)   & 0.00624 \\
$B(E1; 3^-_{\text{band8}}\rightarrow4^+_{\text{band2}})$    &  0.0033(+7-6)   & 0.00747 \\
$B(E1; 1^-_{\text{band8}}\rightarrow2^+_{\text{band3}})$    &  5.9E-5(18)     & 0.000010 \\
$B(E1; 1^-_{\text{band8}}\rightarrow0^+_{\text{band4}})$    &  0.000117(19)   & 0.000018 \\
$B(E1; 1^-_{\text{band8}}\rightarrow2^+_{\text{band4}})$    &  7.2E-5(28)     & 0.0024 \\
$B(E1; 2^-_{\text{band9}}\rightarrow2^+_{\text{band1}})$    &  1.22E-7(+24-18)& 1.82E-7 \\
$B(E1; 3^-_{\text{band9}}\rightarrow2^+_{\text{band1}})$     &  6.9E-5(9)      & 0.0000007 \\
$B(E1; 3^-_{\text{band9}}\rightarrow4^+_{\text{band1}})$     &  0.00054(+7-6)  & 0.0000043 \\
$B(E1; 2^-_{\text{band9}}\rightarrow2^+_{\text{band2}})$    &  4.7E-8(+9-7)   & 1.44E-7 \\
$B(E1; 3^-_{\text{band9}}\rightarrow2^+_{\text{band2}})$     &  0.00044(6)     & 0.0000063 \\
$B(E1; 3^-_{\text{band9}}\rightarrow4^+_{\text{band2}})$     &  0.00021(3)     & 0.0000050 \\
$B(E1; 2^-_{\text{band9}}\rightarrow2^+_{\text{band3}})$    &  4.3E-6(+9-7)   & 0.00422 \\
$B(E1; 2^-_{\text{band9}}\rightarrow3^+_{\text{band3}})$    &  2.4E-6(+5-4)   & 0.00236 \\
$B(E1; 3^-_{\text{band9}}\rightarrow2^+_{\text{band3}})$     &  0.00136(+20-19)& 0.00162 \\
$B(E1; 3^-_{\text{band9}}\rightarrow3^+_{\text{band3}})$     &  0.0026(10)     & 0.00253 \\
$B(E1; 3^-_{\text{band9}}\rightarrow4^+_{\text{band3}})$     &  0.0019(3)      & 0.00334 \\
$B(E1; 2^-_{\text{band9}}\rightarrow2^+_{\text{band4}})$    &  2.1E-7(+4-3)   & 3.188E-6 \\
$B(E3; 3^-_{\text{band7}}\rightarrow0^+_{\text{band1}})$    &  14.1(19)       & 23.1 \\
\hline
\hline
\end{tabular}
\end{center}
\end{table}

For the negative-parity bands, the lowest $1^-_1$ band (band 7) exhibits intraband transitions similar to those of the ground-state band, with an extremely strong interband $B(E1)$ transition connecting them (see Table \ref{Tab-E1transition}). However, the $1^-_1$ bandhead is predicted at a relatively high energy ($\sim 1$ MeV), which precludes the observation of a parity doublet with the ground-state band. The calculated $B(E3; 3^-_1\to 0^+_1)$ value is 23.1 W.U., compared to the experimental value of 14.1(19) W.U.~\cite{KibediADNDT2002}. Both values are significantly lower than those of typical pear-shaped nuclei like $^{144,146}$Ba in this mass region \cite{BucherPRL2016,BucherPRL2017}. Combined with the softness of the deformation energy surface (DES) along $\beta_{30}$ shown in Fig.~\ref{PES-Sm152}, these observations support the interpretation of the $1^-_1$ band as an octupole-soft band \cite{ButlerRMP1996,ButlerPRL2020}. Band 8, with a $1^-$ bandhead at 1681 keV, has been previously assigned as a $K^\pi=0^-$ octupole excitation built on the $0^+_2$ band, based on its similar E1 decay pattern to the ground-state band as band 7 and energy considerations \cite{GarrettPRL2009}. Our calculations reproduce the general structure of band 8 and its interband $B(E1)$ transitions to the $0^+_2$ band reasonably well (Table \ref{Tab-E1transition}). Furthermore, the predicted $B(E2)$ values for intraband transitions are similar to those of the $0^+_2$ band, consistent with their analogous probability density distributions along quadrupole deformations (Figs. \ref{Wav-0+1} and \ref{Wav-1-1}). A significant discrepancy, however, is that the $B(E1)$ values for transitions from band 8 to the ground-state band are overestimated by about two orders of magnitude. One possible explanation is the mixing of a $K^\pi=1^-$ component into this band, given its proximity to the $K^\pi=1^-$ band (band 11). Such mixing is not accounted for in our present model, which excludes K=1 components.

For the first $K^\pi=2^-$ band (band 9), the model predictions show good agreement with experimental data, accurately reproducing both the band structure and the strong $E1$ transitions to the $\gamma$-band. The calculations also predict a second band with similar spins and parity ($2^-_2, 3^-_4, 4^-_2, \cdots$) at higher excitation energies, which exhibits weaker $E1$ and $E2$ transitions to the $\gamma$-band. Analysis of the probability density distributions in Fig.~\ref{Wav-1-1} reveals the distinct structural origins of these bands: the first $K^\pi=2^-$ band is identified as an axial octupole excitation built on the $\gamma$-band, whereas the second is a vibrational excitation along the triaxial octupole deformation $\beta_{32}$. To facilitate the experimental identification of this predicted triaxial octupole band, we provide a comprehensive list of its relevant $B(E\lambda)$ $(\lambda=1,2,3)$ values in Appendix \ref{sec:APPTab}. Furthermore, the model predicts a $K^\pi=4^-$ band with a bandhead $4^-_3$, which is a potential candidate for the experimentally observed sequence of $5^-, 6^-, 7^-, \cdots$ states (band 10). A definitive assignment, however, would require measurements of transition rates from this band to known states. Finally, the present model does not describe the $K^\pi=1^-$ band (band 11), as the necessary $\beta_{31}$ degree of freedom is not included in the current framework.

\begin{figure}[ht]
  \centering{\includegraphics[width=0.46\textwidth]{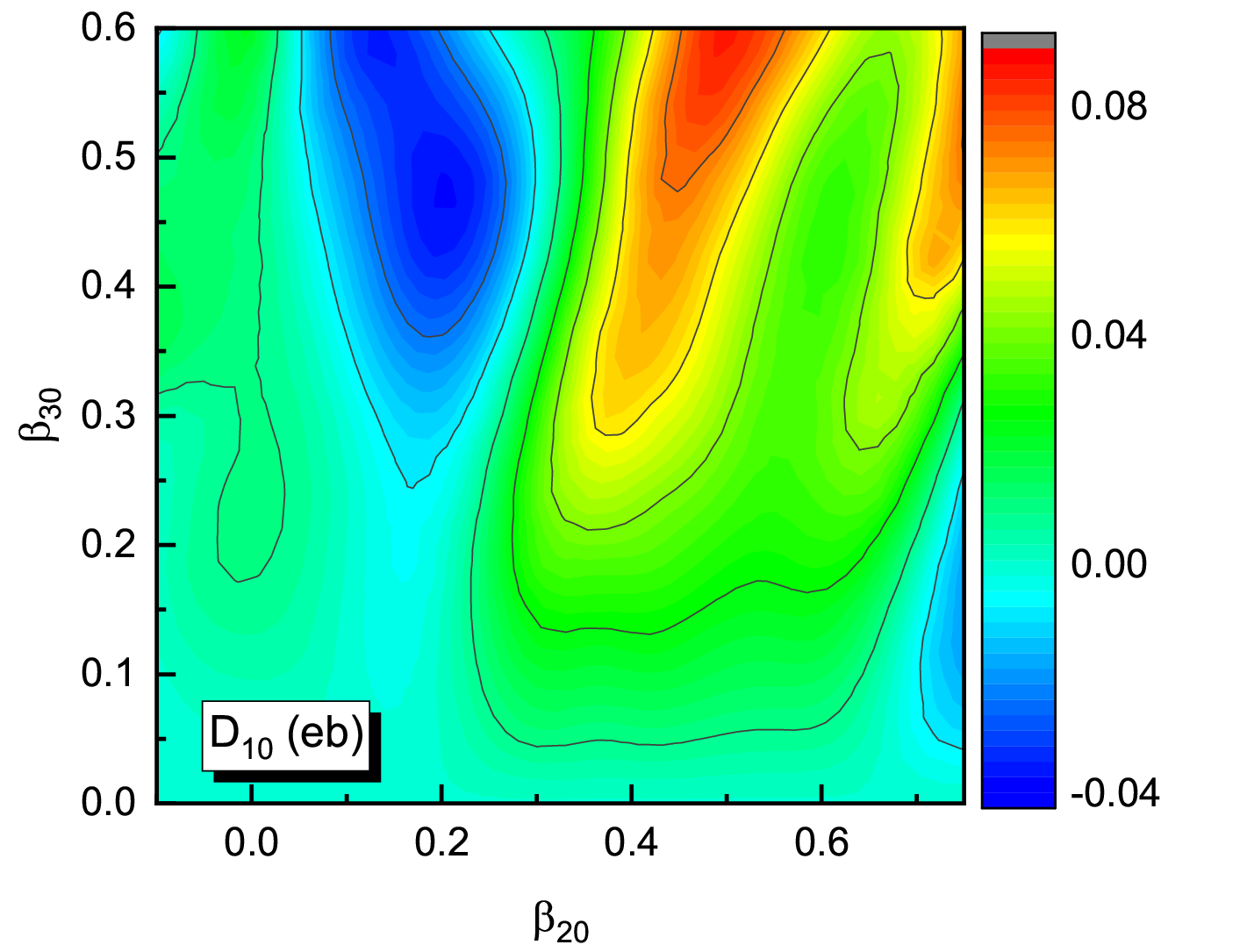}}
  \caption{\label{Fig:D10}(Color online) Intrinsic electric dipole moment $D_{10}$ (in units of $eb$) of $^{152}$Sm in the $(\beta_{20}, \beta_{30})$ plane. The remaining collective degrees of freedom are fixed at the values that correspond to the global minimum. }
\end{figure}

The microscopic TQOCH model successfully reproduces most of the experimental intraband $B(E2)$ values (Fig.~\ref{Spectra-Sm152}) and interband $\rho^2(E0)$, $B(E2)$ and $B(E3)$ values (Tables \ref{Tab-E2transition} and \ref{Tab-E1transition}). This agreement is notable given that the theoretical values span several orders of magnitude and the model uses no effective charges. For strong $E1$ transitions, the theoretical results agree with the data within a factor of approximately two. This is a reasonable description, especially considering that E1 transitions are highly sensitive not only to collectivity but also to the underlying shell structure \cite{ButlerRMP1996}. This shell-structure sensitivity is illustrated in Fig.~\ref{Fig:D10}, which shows the intrinsic electric dipole moment $D_{10}$ across the $(\beta_{20}, \beta_{30})$ deformation plane. The value of $D_{10}$ varies rapidly with both quadrupole and octupole deformations, even changing sign at $\beta_{20}\approx0.25$ due to shell effects. Consequently, an accurate description of E1 transitions requires a precise depiction of this fine shell structure evolution. A further limitation of the present calculation is that it includes only the axially symmetric component of the intrinsic dipole moment. A more complete treatment, which includes $\beta_{31}$ deformation, is necessary and is currently in progress.

To further elucidate the structure of the low-lying bands in $^{152}$Sm, we plot the TQOCH-calculated probability density distributions for the bandheads across four deformation planes: ($\beta_{20},\beta_{22}$), $(\beta_{20},\beta_{30})$, $(\beta_{30},\beta_{32})$, and ($\beta_{30},\beta_{22}$). These two-dimensional distributions in the $(\beta_{\mu\nu}, \beta_{\mu^\prime\nu^\prime})$ plane are generated by integrating the full probability density over all other degrees of freedom: 

\begin{equation}
\mathcal{\rho}_{\alpha}^{{I}\pi}(\beta_{\mu\nu}, \beta_{\mu^\prime\nu^\prime})=\int{\rm d}\beta_{ij}{\rm d}\beta_{i^\prime j^\prime} \sum\limits_{K\in\text{even}}\left|\psi^{I\pi}_{\alpha{K}}\left(\beta_{\mu\nu}, \beta_{\mu^\prime\nu^\prime}, \beta_{ij}, \beta_{i^\prime j^\prime}\right)\right|^2.
\end{equation}

For the positive-parity bandheads in Fig.~\ref{Wav-0+1}, the probability density distributions are broad across the octupole deformation region ($\beta_{30}\in[-0.2, 0.2]$), reflecting a soft potential energy surface in this degree of freedom. The first three $0^+$ states represent the typical sequence of the ground state, $\beta$-vibration, and a second $\beta$-vibration, characterized by zero, one, and two phonons in the $\beta_{20}$ deformation, respectively. It is noteworthy that the $0^+_3$ state has been identified as a pairing isomer in some experimental studies \cite{KulpPRC2005,HindsPL1965,DebenhamNPA1972}. This specific excitation arises from the dramatically different pairing gaps for oblate orbitals (such as the up-sloping $\frac{11}{2}[505]_\nu$ orbital) compared to prolate orbitals. Its key experimental signature is the strikingly different population of the $0^+$ bandhead in $(p,t)$ versus $(t,p)$ reactions. However, if the $0^+_3$ state is indeed a pairing isomer, the strong interband transitions to the $0^+_2$ band presents a challenge to this interpretation and requires further understanding. The states $0^+_4$, $2^+_3$, and $4^+_6$ all show $\gamma$-vibrational character, with peaks at nonzero $\beta_{22}$. The $2^+_3$ state is a pure $K=2$ $\gamma$-vibration, and the $0^+_4$ state is an $n_\gamma=1$ $\gamma$-vibration, based on its nodal structure and energy. The $4^+_6$ state is a second $\gamma$-vibration with strong $K=0$ and $K=4$ mixing (cf. Fig.~ \ref{Nk-Sm152}).

Figure~\ref{Wav-1-1} shows that the probability density distributions for the lowest-lying negative-parity band ($1^-_1$ and $3^-_1$) are similar to that of the ground state, differing primarily by a node along the $\beta_{30}$ axis. The second $1^-$ state displays nodes along both the $\beta_{20}$ and $\beta_{30}$ directions, identifying it as an octupole excitation built upon the $\beta$-vibrational state ($0^+_2$). The first $K^\pi=2^-$ state, $2^-_1$, exhibits a nonzero $\beta_{22}$ peak and a node along $\beta_{30}$, suggesting an octupole excitation based on the $\gamma$-band. In contrast, the second $K^\pi=2^-$ state, $2^-_2$, has a peak at $\beta_{32}\approx 0.08$, which is characteristic of a triaxial octupole vibration. Finally, the probability density distributions of the $4^-_3$ state ($K = 4$) are very similar to those of the $2^-_1$ state, implying it is an octupole excitation built upon the second $\gamma$-vibration.


Finally, we discuss the potential existence of tetrahedral shapes and their associated rotational bands in the rare-earth region. In 2006, J. Dudek {\it et al.} \cite{DudekPRL2006} used macroscopic-microscopic calculations of multidimensional potential energy surfaces to propose that nuclear shapes with tetrahedral ($T_d$) or octahedral ($O_h$) symmetries could exist just a few hundred keV above the ground state in specific rare-earth nuclei near $^{156}$Gd and $^{160}$Yb. Subsequently, Tagami {\it et al.} \cite{TagamiPRC2013} applied group representation theory to predict the spin-parity sequence of rotational bands built on an $I^\pi = 0^+$ bandhead in a $T_d$-symmetric nucleus: $0^+, 3^-, 4^+, (6^+, 6^-), 7^-, 8^+, (9^+, 9^-), (10^+, 10^-), \ldots$, where $(I^+, I^-)$ denotes a parity doublet.

Between 2010 and 2017, experimental searches for the low-lying negative-parity bands characteristic of tetrahedral symmetry were conducted in candidate rare-earth \cite{BarkPRL2010,JentschelPRL2010,DudekPRL2006,DoanPRC2010,HartleyPRC2017,DobrowolskiIJMPE2011} and actinide nuclei \cite{NtshangasePRC2010}, but none proved successful. More recently, however, S. Basak {\it et al.} \cite{BasakPRC2025}  reported evidence for a new tetrahedral band in $^{152}$Sm, with a $3^-$ bandhead at 1933.5 KeV. The observed mixed-parity sequence, which lacks $E2$ transitions, is consistent with the spectroscopic criteria for a band with tetrahedral symmetry. Nevertheless, no reduced transition strengths for this band have been determined to confirm this assignment.

The present TQOCH model, based on a fully microscopic four-dimensional energy surface, finds no evidence for a local tetrahedral minimum and consequently predicts no tetrahedral rotational band in $^{152}$Sm. To confirm the experimental assignments, direct measurements of the key interband B(E1) and intraband B(E3) values are essential. Furthermore, the theoretical picture requires consolidation through calculations with different density functionals to test the robustness of this prediction.


\section{Summary and outlook}\label{Sec:IV}

In summary, we have developed a microscopic Triaxial Quadrupole-Octupole Collective Hamiltonian (TQOCH) to model the low-lying positive- and negative-parity states of nuclei. The TQOCH framework simultaneously describes axial and triaxial quadrupole vibrations, octupole vibrations, rotations, and their couplings. Its dynamics are determined entirely by a multidimensionally constrained Covariant Density Functional Theory (CDFT) calculation, which maps the potential energy surface across the quadrupole ($\beta_{20}, \beta_{22}$) and octupole ($\beta_{30}, \beta_{32}$) deformation parameters. From this, single-nucleon properties are used to calculate the microscopic inputs for the collective Hamiltonian: ten mass parameters  $B_{\mu\nu,\mu^\prime\nu^\prime}$, three moments of inertia ${\cal I}_k$, and the collective potential (including zero-point energy corrections). Finally, diagonalizing the Hamiltonian yields excitation energies, collective wave functions, and electromagnetic transition rates for the low-lying states.

The new model was extensively tested on the transitional nucleus $^{152}$Sm, which has rich spectroscopic data for both positive- and negative-parity bands. Calculations of deformation energy surfaces, excitation spectra, electric multipole transitions, and probability density distributions demonstrate that the TQOCH accurately reproduces the band structure and transitions for all positive-parity bands and the $K^\pi=0^-, 2^-$ negative-parity bands. Our analysis reveals two distinct excitation modes for the $K^\pi=2^-$ band: an axially symmetric octupole excitation built on the $\gamma$-band and a triaxial octupole vibration, identified by their unique $\beta_{22}$ and $\beta_{32}$ probability distributions. Furthermore, we predict a $K^\pi=4^-$ band (bandhead $4^-_3$) that could correspond to the experimentally observed $5^-, 6^-, 7^-, \cdots$ negative-parity sequence. Confirming this assignment will require future measurements of the relevant transition rates.

A limitation of the current TQOCH is its exclusion of the $\beta_{31}$ deformation, which prevents the description of the $K^\pi=1^-$ band and is likely linked to the overestimation of certain $E1$ transitions. The inclusion of this degree of freedom is therefore a necessary next step. Fortunately, our foundational CDFT calculations already incorporate $\beta_{31}$ and $\beta_{33}$  deformations \cite{ZhaoPRC2024a}, providing the required microscopic input. An extension of the TQOCH to include these degrees of freedom is now in progress.

\begin{acknowledgments}
This work has been supported in part by the National Natural Science Foundation of China under Grants No. 12005109, No. 12375126, the PHD Foundation of Chongqing Normal University (No. 23XLB010), the Science and Technology Research Program of Chongqing Municipal Education Commission (No. KJQN202300509), by the project “Implementation of cutting-edge research and its application as part of the Scientific Center of Excellence for Quantum and Complex Systems, and Representations of Lie Algebras“, Grant No. PK.1.1.10.0004, co-financed by the European Union through the European Regional Development Fund - Competitiveness and Cohesion Programme 2021-2027, and by the Croatian Science Foundation under the project Relativistic Nuclear Many-Body Theory in the Multimessenger Observation Era (IP-2022-10-7773).
\end{acknowledgments}

\appendix

\section{Theoretical $B(E1)$, $B(E2)$, and $B(E3)$ values for transitions in the $K^\pi=2^-$ and $4^-$ bands}\label{sec:APPTab}
\begin{table}[htbp]
\tabcolsep=4pt   
\begin{center}
\caption{\label{Tab-E1-App} TQOCH $B(E1)$ values for transitions from the $2^-_2$ band (band 12), and the $4^-_3$ band (band 10) of $^{152}$Sm.}
\begin{tabular}{ccc|ccc}
\hline\hline
$I^{\pi}_i$ & $I^{\pi}_f$  &  TQOCH (W.U.)  & $I^{\pi}_i$ & $I^{\pi}_f$  &  TQOCH (W.U.)\\
\hline
$2^-_{\text{band12}}$   &   $2^+_{\text{band3}}$  &  0.00241  &  $9^-_{\text{band12}}$   &   $10^+_{\text{band5}}$ &  4.79E-5 \\
$2^-_{\text{band12}}$   &   $2^+_{\text{band5}}$  &  2.74E-5  &  $9^-_{\text{band12}}$   &   $10^+_{\text{band6}}$ &  5.38E-5 \\
$2^-_{\text{band12}}$   &   $3^+_{\text{band3}}$  &  0.00126  &  $10^-_{\text{band12}}$  &   $9^+_{\text{band3}}$  &  2.71E-4 \\
$3^-_{\text{band12}}$   &   $2^+_{\text{band2}}$  &  2.91E-5  &  $10^-_{\text{band12}}$  &   $10^+_{\text{band3}}$ &  2.06E-5 \\
$3^-_{\text{band12}}$   &   $2^+_{\text{band3}}$  &  7.76E-4  &  $10^-_{\text{band12}}$  &   $10^+_{\text{band5}}$ &  2.03E-6 \\
$3^-_{\text{band12}}$   &   $2^+_{\text{band5}}$  &  8.81E-6  &  $10^-_{\text{band12}}$  &   $10^+_{\text{band6}}$ &  1.62E-5 \\
$3^-_{\text{band12}}$   &   $3^+_{\text{band3}}$  &  0.00114  &  $4^-_{\text{band10}}$   &   $3^+_{\text{band3}}$  &  7.27E-6 \\
$3^-_{\text{band12}}$   &   $4^+_{\text{band1}}$  &  1.94E-6  &  $4^-_{\text{band10}}$   &   $4^+_{\text{band5}}$  &  6.24E-4 \\
$3^-_{\text{band12}}$   &   $4^+_{\text{band2}}$  &  4.80E-5  &  $4^-_{\text{band10}}$   &   $4^+_{\text{band6}}$  &  6.42E-4 \\
$3^-_{\text{band12}}$   &   $4^+_{\text{band3}}$  &  0.00143  &  $4^-_{\text{band10}}$   &   $5^+_{\text{band3}}$  &  2.63E-6 \\
$3^-_{\text{band12}}$   &   $4^+_{\text{band5}}$  &  8.13E-5  &  $5^-_{\text{band10}}$   &   $4^+_{\text{band2}}$  &  9.76E-5 \\
$3^-_{\text{band12}}$   &   $4^+_{\text{band6}}$  &  1.42E-5  &  $5^-_{\text{band10}}$   &   $4^+_{\text{band3}}$  &  1.64E-5 \\
$4^-_{\text{band12}}$   &   $3^+_{\text{band3}}$  &  9.70E-4  &  $5^-_{\text{band10}}$   &   $4^+_{\text{band4}}$  &  5.50E-5 \\
$4^-_{\text{band12}}$   &   $4^+_{\text{band3}}$  &  5.72E-4  &  $5^-_{\text{band10}}$   &   $4^+_{\text{band5}}$  &  1.48E-4 \\
$4^-_{\text{band12}}$   &   $4^+_{\text{band5}}$  &  2.44E-5  &  $5^-_{\text{band10}}$   &   $4^+_{\text{band6}}$  &  2.30E-4 \\
$4^-_{\text{band12}}$   &   $4^+_{\text{band6}}$  &  8.74E-6  &  $5^-_{\text{band10}}$   &   $4^+_{\text{band1}}$  &  1.42E-6 \\
$4^-_{\text{band12}}$   &   $5^+_{\text{band3}}$  &  0.00142  &  $5^-_{\text{band10}}$   &   $4^+_{\text{band2}}$  &  1.26E-4 \\
$5^-_{\text{band12}}$   &   $4^+_{\text{band1}}$  &  2.69E-6  &  $5^-_{\text{band10}}$   &   $4^+_{\text{band3}}$  &  7.79E-6 \\
$5^-_{\text{band12}}$   &   $4^+_{\text{band2}}$  &  1.07E-4  &  $5^-_{\text{band10}}$   &   $4^+_{\text{band4}}$  &  5.59E-5 \\
$5^-_{\text{band12}}$   &   $4^+_{\text{band3}}$  &  9.88E-4  &  $5^-_{\text{band10}}$   &   $4^+_{\text{band5}}$  &  7.09E-5 \\
$5^-_{\text{band12}}$   &   $4^+_{\text{band5}}$  &  4.36E-5  &  $5^-_{\text{band10}}$   &   $4^+_{\text{band6}}$  &  0.00100 \\
$5^-_{\text{band12}}$   &   $4^+_{\text{band6}}$  &  1.19E-5  &  $6^-_{\text{band10}}$   &   $5^+_{\text{band3}}$  &  3.19E-5 \\
$5^-_{\text{band12}}$   &   $5^+_{\text{band3}}$  &  3.77E-4  &  $6^-_{\text{band10}}$   &   $6^+_{\text{band5}}$  &  1.86E-4 \\
$5^-_{\text{band12}}$   &   $6^+_{\text{band1}}$  &  4.67E-6  &  $6^-_{\text{band10}}$   &   $6^+_{\text{band6}}$  &  0.00130 \\
$5^-_{\text{band12}}$   &   $6^+_{\text{band2}}$  &  1.47E-4  &  $6^-_{\text{band10}}$   &   $7^+_{\text{band3}}$  &  1.25E-5 \\
$5^-_{\text{band12}}$   &   $6^+_{\text{band3}}$  &  0.00122  &  $7^-_{\text{band10}}$   &   $6^+_{\text{band1}}$  &  3.71E-5 \\
$5^-_{\text{band12}}$   &   $6^+_{\text{band5}}$  &  1.50E-4  &  $7^-_{\text{band10}}$   &   $6^+_{\text{band2}}$  &  2.29E-4 \\
$6^-_{\text{band12}}$   &   $5^+_{\text{band3}}$  &  8.70E-4  &  $7^-_{\text{band10}}$   &   $6^+_{\text{band3}}$  &  3.78E-5 \\
$6^-_{\text{band12}}$   &   $6^+_{\text{band3}}$  &  1.89E-4  &  $7^-_{\text{band10}}$   &   $6^+_{\text{band4}}$  &  3.66E-5 \\
$6^-_{\text{band12}}$   &   $6^+_{\text{band5}}$  &  1.74E-5  &  $7^-_{\text{band10}}$   &   $6^+_{\text{band5}}$  &  1.21E-4 \\
$6^-_{\text{band12}}$   &   $6^+_{\text{band6}}$  &  2.56E-6  &  $7^-_{\text{band10}}$   &   $6^+_{\text{band6}}$  &  0.00141 \\
$6^-_{\text{band12}}$   &   $7^+_{\text{band3}}$  &  0.00106  &  $7^-_{\text{band10}}$   &   $7^+_{\text{band3}}$  &  4.58E-4 \\
$7^-_{\text{band12}}$   &   $6^+_{\text{band1}}$  &  5.26E-6  &  $7^-_{\text{band10}}$   &   $8^+_{\text{band1}}$  &  3.02E-5 \\
$7^-_{\text{band12}}$   &   $6^+_{\text{band2}}$  &  2.52E-4  &  $7^-_{\text{band10}}$   &   $8^+_{\text{band2}}$  &  3.05E-4 \\
$7^-_{\text{band12}}$   &   $6^+_{\text{band3}}$  &  7.73E-4  &  $7^-_{\text{band10}}$   &   $8^+_{\text{band3}}$  &  1.90E-5 \\
$7^-_{\text{band12}}$   &   $6^+_{\text{band4}}$  &  3.92E-6  &  $7^-_{\text{band10}}$   &   $8^+_{\text{band4}}$  &  2.69E-5 \\
$7^-_{\text{band12}}$   &   $6^+_{\text{band5}}$  &  7.00E-5  &  $7^-_{\text{band10}}$   &   $8^+_{\text{band5}}$  &  1.33E-4 \\
$7^-_{\text{band12}}$   &   $6^+_{\text{band6}}$  &  2.13E-6  &  $7^-_{\text{band10}}$   &   $8^+_{\text{band6}}$  &  0.00185 \\
$7^-_{\text{band12}}$   &   $7^+_{\text{band3}}$  &  1.50E-4  &  $8^-_{\text{band10}}$   &   $7^+_{\text{band3}}$  &  5.58E-5 \\
$7^-_{\text{band12}}$   &   $8^+_{\text{band1}}$  &  5.13E-6  &  $8^-_{\text{band10}}$   &   $8^+_{\text{band3}}$  &  2.33E-6 \\
$7^-_{\text{band12}}$   &   $8^+_{\text{band2}}$  &  3.22E-4  &  $8^-_{\text{band10}}$   &   $8^+_{\text{band5}}$  &  1.21E-4 \\
$7^-_{\text{band12}}$   &   $8^+_{\text{band3}}$  &  8.61E-4  &  $8^-_{\text{band10}}$   &   $8^+_{\text{band6}}$  &  0.00110 \\
$7^-_{\text{band12}}$   &   $8^+_{\text{band5}}$  &  1.21E-4  &  $8^-_{\text{band10}}$   &   $9^+_{\text{band3}}$  &  2.18E-5 \\
$7^-_{\text{band12}}$   &   $8^+_{\text{band6}}$  &  8.34E-6  &  $9^-_{\text{band10}}$   &   $8^+_{\text{band1}}$  &  5.54E-5 \\
$8^-_{\text{band12}}$   &   $7^+_{\text{band3}}$  &  5.77E-4  &  $9^-_{\text{band10}}$   &   $8^+_{\text{band2}}$  &  1.72E-4 \\
$8^-_{\text{band12}}$   &   $8^+_{\text{band3}}$  &  6.66E-5  &  $9^-_{\text{band10}}$   &   $8^+_{\text{band3}}$  &  4.72E-5 \\
$8^-_{\text{band12}}$   &   $8^+_{\text{band5}}$  &  5.84E-6  &  $9^-_{\text{band10}}$   &   $8^+_{\text{band4}}$  &  1.82E-5 \\
$8^-_{\text{band12}}$   &   $8^+_{\text{band6}}$  &  8.78E-6  &  $9^-_{\text{band10}}$   &   $8^+_{\text{band5}}$  &  1.17E-4 \\
$8^-_{\text{band12}}$   &   $9^+_{\text{band3}}$  &  6.51E-4  &  $9^-_{\text{band10}}$   &   $8^+_{\text{band6}}$  &  0.00237 \\
$9^-_{\text{band12}}$   &   $8^+_{\text{band1}}$  &  6.84E-6  &  $9^-_{\text{band10}}$   &   $9^+_{\text{band3}}$  &  1.40E-5 \\
$9^-_{\text{band12}}$   &   $8^+_{\text{band2}}$  &  6.10E-4  &  $9^-_{\text{band10}}$   &   $10^+_{\text{band1}}$ &  4.76E-5 \\
$9^-_{\text{band12}}$   &   $8^+_{\text{band3}}$  &  4.56E-4  &  $9^-_{\text{band10}}$   &   $10^+_{\text{band2}}$ &  2.34E-4 \\
$9^-_{\text{band12}}$   &   $8^+_{\text{band4}}$  &  4.51E-4  &  $9^-_{\text{band10}}$   &   $10^+_{\text{band3}}$ &  2.61E-5 \\
$9^-_{\text{band12}}$   &   $8^+_{\text{band5}}$  &  2.59E-5  &  $9^-_{\text{band10}}$   &   $10^+_{\text{band4}}$ &  1.19E-5 \\
$9^-_{\text{band12}}$   &   $8^+_{\text{band6}}$  &  1.21E-5  &  $9^-_{\text{band10}}$   &   $10^+_{\text{band5}}$ &  1.10E-5 \\
$9^-_{\text{band12}}$   &   $9^+_{\text{band3}}$  &  5.64E-5  &  $9^-_{\text{band10}}$   &   $10^+_{\text{band6}}$ &  0.00122 \\
$9^-_{\text{band12}}$   &   $10^+_{\text{band1}}$ &  2.38E-6  &  $10^-_{\text{band10}}$  &   $9^+_{\text{band3}}$  &  6.94E-5 \\
$9^-_{\text{band12}}$   &   $10^+_{\text{band2}}$ &  7.55E-4  &  $10^-_{\text{band10}}$  &   $10^+_{\text{band3}}$ &  5.39E-6 \\
$9^-_{\text{band12}}$   &   $10^+_{\text{band3}}$ &  4.84E-4  &  $10^-_{\text{band10}}$  &   $10^+_{\text{band5}}$ &  2.19E-5 \\
$9^-_{\text{band12}}$   &   $10^+_{\text{band4}}$ &  2.04E-5  &  $10^-_{\text{band10}}$  &   $10^+_{\text{band6}}$ &  4.49E-4 \\
\hline
\hline
\end{tabular}
\end{center}
\end{table}

\begin{table}[htbp]
\tabcolsep=4pt   
\begin{center}
\caption{\label{Tab-E2-APP-2}  Theoretical $B(E2)$ values for transitions from the $4^-_3$ band (band 10) of $^{152}$Sm.}
\begin{tabular}{ccc|ccc}
\hline\hline
$I^{\pi}_i$ & $I^{\pi}_f$  &  TQOCH (W.U.)  & $I^{\pi}_i$ & $I^{\pi}_f$  &  TQOCH (W.U.) \\
\hline
$4^-_{\text{band10}}$   &   $2^-_{\text{band9}}$  &  17.4     &  $7^-_{\text{band10}}$   &   $6^-_{\text{band10}}$ &  134    \\
$4^-_{\text{band10}}$   &   $2^-_{\text{band12}}$ &  1.51     &  $7^-_{\text{band10}}$   &   $7^-_{\text{band7}}$  &  0.0494 \\
$4^-_{\text{band10}}$   &   $3^-_{\text{band8}}$  &  0.0253   &  $7^-_{\text{band10}}$   &   $7^-_{\text{band8}}$  &  0.0724 \\
$4^-_{\text{band10}}$   &   $3^-_{\text{band9}}$  &  13.5     &  $7^-_{\text{band10}}$   &   $7^-_{\text{band9}}$  &  8.69   \\
$4^-_{\text{band10}}$   &   $3^-_{\text{band12}}$ &  1.57     &  $7^-_{\text{band10}}$   &   $7^-_{\text{band12}}$ &  0.583  \\
$4^-_{\text{band10}}$   &   $4^-_{\text{band9}}$  &  6.24     &  $7^-_{\text{band10}}$   &   $8^-_{\text{band9}}$  &  8.21   \\
$4^-_{\text{band10}}$   &   $4^-_{\text{band12}}$ &  0.912    &  $7^-_{\text{band10}}$   &   $9^-_{\text{band7}}$  &  0.0164 \\
$4^-_{\text{band10}}$   &   $5^-_{\text{band9}}$  &  1.26     &  $7^-_{\text{band10}}$   &   $9^-_{\text{band8}}$  &  0.797  \\
$4^-_{\text{band10}}$   &   $5^-_{\text{band12}}$ &  0.0818   &  $7^-_{\text{band10}}$   &   $9^-_{\text{band9}}$  &  0.0139 \\
$4^-_{\text{band10}}$   &   $6^-_{\text{band9}}$  &  0.873    &  $8^-_{\text{band10}}$   &   $6^-_{\text{band9}}$  &  2.34   \\
$5^-_{\text{band10}}$   &   $3^-_{\text{band8}}$  &  0.0177   &  $8^-_{\text{band10}}$   &   $6^-_{\text{band12}}$ &  0.897  \\
$5^-_{\text{band10}}$   &   $3^-_{\text{band9}}$  &  8.70     &  $8^-_{\text{band10}}$   &   $6^-_{\text{band10}}$ &  85.6   \\
$5^-_{\text{band10}}$   &   $3^-_{\text{band12}}$ &  0.668    &  $8^-_{\text{band10}}$   &   $7^-_{\text{band7}}$  &  0.0206 \\
$5^-_{\text{band10}}$   &   $4^-_{\text{band9}}$  &  14.9     &  $8^-_{\text{band10}}$   &   $7^-_{\text{band8}}$  &  0.0440 \\
$5^-_{\text{band10}}$   &   $4^-_{\text{band12}}$ &  1.12     &  $8^-_{\text{band10}}$   &   $7^-_{\text{band9}}$  &  9.21   \\
$5^-_{\text{band10}}$   &   $4^-_{\text{band10}}$ &  131      &  $8^-_{\text{band10}}$   &   $7^-_{\text{band12}}$ &  0.606  \\
$5^-_{\text{band10}}$   &   $5^-_{\text{band8}}$  &  0.282    &  $8^-_{\text{band10}}$   &   $7^-_{\text{band10}}$ &  117    \\
$5^-_{\text{band10}}$   &   $5^-_{\text{band9}}$  &  8.23     &  $8^-_{\text{band10}}$   &   $8^-_{\text{band9}}$  &  13.4   \\
$5^-_{\text{band10}}$   &   $5^-_{\text{band12}}$ &  1.09     &  $8^-_{\text{band10}}$   &   $8^-_{\text{band12}}$ &  0.757  \\
$5^-_{\text{band10}}$   &   $6^-_{\text{band9}}$  &  3.86     &  $8^-_{\text{band10}}$   &   $9^-_{\text{band7}}$  &  0.0238 \\
$5^-_{\text{band10}}$   &   $7^-_{\text{band8}}$  &  0.268    &  $8^-_{\text{band10}}$   &   $9^-_{\text{band8}}$  &  0.0373 \\
$5^-_{\text{band10}}$   &   $7^-_{\text{band9}}$  &  0.0111   &  $8^-_{\text{band10}}$   &   $9^-_{\text{band9}}$  &  5.34   \\
$6^-_{\text{band10}}$   &   $4^-_{\text{band9}}$  &  5.38     &  $8^-_{\text{band10}}$   &   $9^-_{\text{band12}}$ &  0.246  \\
$6^-_{\text{band10}}$   &   $4^-_{\text{band12}}$ &  0.709    &  $8^-_{\text{band10}}$   &   $10^-_{\text{band9}}$ &  0.348  \\
$6^-_{\text{band10}}$   &   $4^-_{\text{band10}}$ &  30.4     &  $9^-_{\text{band10}}$   &   $7^-_{\text{band8}}$  &  0.160  \\
$6^-_{\text{band10}}$   &   $5^-_{\text{band7}}$  &  0.0151   &  $9^-_{\text{band10}}$   &   $7^-_{\text{band9}}$  &  1.27   \\
$6^-_{\text{band10}}$   &   $5^-_{\text{band8}}$  &  0.0496   &  $9^-_{\text{band10}}$   &   $7^-_{\text{band12}}$ &  0.671  \\
$6^-_{\text{band10}}$   &   $5^-_{\text{band9}}$  &  12.6     &  $9^-_{\text{band10}}$   &   $7^-_{\text{band10}}$ &  112    \\
$6^-_{\text{band10}}$   &   $5^-_{\text{band12}}$ &  0.780    &  $9^-_{\text{band10}}$   &   $8^-_{\text{band9}}$  &  9.38   \\
$6^-_{\text{band10}}$   &   $5^-_{\text{band10}}$ &  146      &  $9^-_{\text{band10}}$   &   $8^-_{\text{band12}}$ &  0.815  \\
$6^-_{\text{band10}}$   &   $6^-_{\text{band9}}$  &  12.0     &  $9^-_{\text{band10}}$   &   $8^-_{\text{band10}}$ &  104    \\
$6^-_{\text{band10}}$   &   $6^-_{\text{band12}}$ &  0.940    &  $9^-_{\text{band10}}$   &   $10^-_{\text{band9}}$ &  11.2   \\
$6^-_{\text{band10}}$   &   $7^-_{\text{band7}}$  &  0.0131   &  $10^-_{\text{band10}}$  &   $8^-_{\text{band9}}$  & 1.17    \\
$6^-_{\text{band10}}$   &   $7^-_{\text{band8}}$  &  0.0251   &  $10^-_{\text{band10}}$  &   $8^-_{\text{band12}}$ & 1.21    \\
$6^-_{\text{band10}}$   &   $7^-_{\text{band9}}$  &  3.80     &  $10^-_{\text{band10}}$  &   $8^-_{\text{band10}}$ & 126     \\
$6^-_{\text{band10}}$   &   $7^-_{\text{band12}}$ &  0.177    &  $10^-_{\text{band10}}$  &   $9^-_{\text{band7}}$  & 0.0222  \\
$6^-_{\text{band10}}$   &   $8^-_{\text{band9}}$  &  0.217    &  $10^-_{\text{band10}}$  &   $9^-_{\text{band8}}$  & 0.0316  \\
$7^-_{\text{band10}}$   &   $5^-_{\text{band8}}$  &  0.220    &  $10^-_{\text{band10}}$  &   $9^-_{\text{band9}}$  & 6.71    \\
$7^-_{\text{band10}}$   &   $5^-_{\text{band9}}$  &  2.97     &  $10^-_{\text{band10}}$  &   $9^-_{\text{band12}}$ & 0.582   \\
$7^-_{\text{band10}}$   &   $5^-_{\text{band12}}$ &  0.587    &  $10^-_{\text{band10}}$  &   $9^-_{\text{band10}}$ & 87.3    \\
$7^-_{\text{band10}}$   &   $5^-_{\text{band10}}$ &  55.2     &  $10^-_{\text{band10}}$  &   $10^-_{\text{band9}}$ & 13.5    \\
$7^-_{\text{band10}}$   &   $6^-_{\text{band9}}$  &  12.2     &   $10^-_{\text{band10}}$ &   $10^-_{\text{band12}}$& 0.549   \\
$7^-_{\text{band10}}$   &   $6^-_{\text{band12}}$ &  0.757 \\
\hline
\hline
\end{tabular}
\end{center}
\end{table}

\begin{table}[htbp]
\tabcolsep=4pt   
\begin{center}
\caption{\label{Tab-E2-APP-1}  TQOCH $B(E2)$ values for transitions from the $2^-_2$ band (band 12) of $^{152}$Sm.}
\begin{tabular}{ccc|ccc}
\hline\hline
$I^{\pi}_i$ & $I^{\pi}_f$  &  TQOCH (W.U.)  & $I^{\pi}_i$ & $I^{\pi}_f$  &  TQOCH (W.U.)\\
\hline
$2^-_{\text{band12}}$   &   $1^-_{\text{band7}}$  &  0.201    &  $7^-_{\text{band12}}$   &   $5^-_{\text{band7}}$  &  0.185  \\
$2^-_{\text{band12}}$   &   $1^-_{\text{band8}}$  &  1.63     &  $7^-_{\text{band12}}$   &   $5^-_{\text{band9}}$  &  0.0261 \\
$2^-_{\text{band12}}$   &   $2^-_{\text{band9}}$  &  1.37     &  $7^-_{\text{band12}}$   &   $5^-_{\text{band12}}$ &  157    \\
$2^-_{\text{band12}}$   &   $3^-_{\text{band7}}$  &  0.0182   &  $7^-_{\text{band12}}$   &   $6^-_{\text{band9}}$  &  0.0419 \\
$2^-_{\text{band12}}$   &   $3^-_{\text{band8}}$  &  0.508    &  $7^-_{\text{band12}}$   &   $6^-_{\text{band12}}$ &  62.7   \\
$2^-_{\text{band12}}$   &   $3^-_{\text{band9}}$  &  3.62     &  $7^-_{\text{band12}}$   &   $6^-_{\text{band10}}$ &  0.153  \\
$2^-_{\text{band12}}$   &   $4^-_{\text{band9}}$  &  2.24     &  $7^-_{\text{band12}}$   &   $7^-_{\text{band7}}$  &  0.421  \\
$3^-_{\text{band12}}$   &   $1^-_{\text{band7}}$  &  0.122    &  $7^-_{\text{band12}}$   &   $7^-_{\text{band8}}$  &  1.89   \\
$3^-_{\text{band12}}$   &   $1^-_{\text{band8}}$  &  0.247    &  $7^-_{\text{band12}}$   &   $7^-_{\text{band9}}$  &  1.96   \\
$3^-_{\text{band12}}$   &   $2^-_{\text{band9}}$  &  0.696    &  $7^-_{\text{band12}}$   &   $8^-_{\text{band9}}$  &  2.07   \\
$3^-_{\text{band12}}$   &   $2^-_{\text{band12}}$ &  181      &  $7^-_{\text{band12}}$   &   $9^-_{\text{band8}}$  &  0.124  \\
$3^-_{\text{band12}}$   &   $3^-_{\text{band7}}$  &  0.162    &  $8^-_{\text{band12}}$   &   $6^-_{\text{band9}}$  &  0.0945 \\
$3^-_{\text{band12}}$   &   $3^-_{\text{band8}}$  &  1.87     &  $8^-_{\text{band12}}$   &   $6^-_{\text{band12}}$ &  164    \\
$3^-_{\text{band12}}$   &   $3^-_{\text{band9}}$  &  0.117    &  $8^-_{\text{band12}}$   &   $6^-_{\text{band10}}$ &  0.0203 \\
$3^-_{\text{band12}}$   &   $4^-_{\text{band9}}$  &  4.37     &  $8^-_{\text{band12}}$   &   $7^-_{\text{band7}}$  &  0.664  \\
$4^-_{\text{band12}}$   &   $2^-_{\text{band9}}$  &  0.0500   &  $8^-_{\text{band12}}$   &   $7^-_{\text{band8}}$  &  0.249  \\
$4^-_{\text{band12}}$   &   $2^-_{\text{band12}}$ &  61.7     &  $8^-_{\text{band12}}$   &   $7^-_{\text{band9}}$  &  0.0914 \\
$4^-_{\text{band12}}$   &   $3^-_{\text{band7}}$  &  0.352    &  $8^-_{\text{band12}}$   &   $7^-_{\text{band12}}$ &  37.3   \\
$4^-_{\text{band12}}$   &   $3^-_{\text{band8}}$  &  0.870    &  $8^-_{\text{band12}}$   &   $7^-_{\text{band10}}$ &  0.494  \\
$4^-_{\text{band12}}$   &   $3^-_{\text{band9}}$  &  0.644    &  $8^-_{\text{band12}}$   &   $8^-_{\text{band9}}$  &  1.74   \\
$4^-_{\text{band12}}$   &   $3^-_{\text{band12}}$ &  128      &  $8^-_{\text{band12}}$   &   $9^-_{\text{band7}}$  &  0.153  \\
$4^-_{\text{band12}}$   &   $4^-_{\text{band9}}$  &  0.220    &  $8^-_{\text{band12}}$   &   $9^-_{\text{band8}}$  &  0.578  \\
$5^-_{\text{band12}}$   &   $3^-_{\text{band7}}$  &  0.157    &  $8^-_{\text{band12}}$   &   $9^-_{\text{band9}}$  &  0.593  \\
$5^-_{\text{band12}}$   &   $3^-_{\text{band8}}$  &  0.0427   &  $8^-_{\text{band12}}$   &   $10^-_{\text{band9}}$ &  2.87   \\
$5^-_{\text{band12}}$   &   $3^-_{\text{band9}}$  &  0.0386   &  $9^-_{\text{band12}}$   &   $7^-_{\text{band7}}$  &  0.182  \\
$5^-_{\text{band12}}$   &   $3^-_{\text{band12}}$ &  106      &  $9^-_{\text{band12}}$   &   $7^-_{\text{band8}}$  &  0.0387 \\
$5^-_{\text{band12}}$   &   $4^-_{\text{band9}}$  &  0.225    &  $9^-_{\text{band12}}$   &   $7^-_{\text{band9}}$  &  0.0209 \\
$5^-_{\text{band12}}$   &   $4^-_{\text{band12}}$ &  102      &  $9^-_{\text{band12}}$   &   $7^-_{\text{band12}}$ &  192    \\
$5^-_{\text{band12}}$   &   $4^-_{\text{band10}}$ &  0.0669   &  $9^-_{\text{band12}}$   &   $7^-_{\text{band10}}$ &  0.0915 \\
$5^-_{\text{band12}}$   &   $5^-_{\text{band7}}$  &  0.281    &  $9^-_{\text{band12}}$   &   $8^-_{\text{band12}}$ &  43.8   \\
$5^-_{\text{band12}}$   &   $5^-_{\text{band8}}$  &  1.92     &  $9^-_{\text{band12}}$   &   $8^-_{\text{band10}}$ &  0.220  \\
$5^-_{\text{band12}}$   &   $5^-_{\text{band9}}$  &  1.47     &  $9^-_{\text{band12}}$   &   $9^-_{\text{band7}}$  &  0.616  \\
$6^-_{\text{band12}}$   &   $4^-_{\text{band9}}$  &  0.0722   &  $9^-_{\text{band12}}$   &   $9^-_{\text{band8}}$  &  2.19   \\
$6^-_{\text{band12}}$   &   $4^-_{\text{band12}}$ &  128      &  $9^-_{\text{band12}}$   &   $9^-_{\text{band9}}$  &  1.72   \\
$6^-_{\text{band12}}$   &   $4^-_{\text{band10}}$ &  0.0120   &  $9^-_{\text{band12}}$   &   $10^-_{\text{band8}}$ &  1.18   \\
$6^-_{\text{band12}}$   &   $5^-_{\text{band7}}$  &  0.517    &  $10^-_{\text{band12}}$  &   $8^-_{\text{band9}}$  &  0.131  \\
$6^-_{\text{band12}}$   &   $5^-_{\text{band8}}$  &  0.477    &  $10^-_{\text{band12}}$  &   $8^-_{\text{band12}}$ &  193    \\
$6^-_{\text{band12}}$   &   $5^-_{\text{band9}}$  &  0.260    &  $10^-_{\text{band12}}$  &   $8^-_{\text{band10}}$ &  0.0255 \\
$6^-_{\text{band12}}$   &   $5^-_{\text{band12}}$ &  65.0     &  $10^-_{\text{band12}}$  &   $9^-_{\text{band7}}$  &  3.10   \\
$6^-_{\text{band12}}$   &   $5^-_{\text{band10}}$ &  0.0415   &  $10^-_{\text{band12}}$  &   $9^-_{\text{band8}}$  &  0.0103 \\
$6^-_{\text{band12}}$   &   $6^-_{\text{band9}}$  &  0.591    &  $10^-_{\text{band12}}$  &   $9^-_{\text{band9}}$  &  22.8   \\
$6^-_{\text{band12}}$   &   $7^-_{\text{band7}}$  &  0.102    &  $10^-_{\text{band12}}$  &   $9^-_{\text{band12}}$ &  1.06   \\
$6^-_{\text{band12}}$   &   $7^-_{\text{band8}}$  &  0.688    &  $10^-_{\text{band12}}$  &   $9^-_{\text{band10}}$ &  0.261  \\
$6^-_{\text{band12}}$   &   $7^-_{\text{band9}}$  &  0.781    &  $10^-_{\text{band12}}$  &   $10^-_{\text{band9}}$ &  0.565  \\
\hline
\hline
\end{tabular}
\end{center}
\end{table}

\begin{table}[htbp]
\tabcolsep=4pt   
\begin{center}
\caption{\label{Tab-E3-APP-1}  TQOCH $B(E3)$ values for transitions from the $2^-_2$ band (band 12), and the $4^-_3$ band (band 10) of $^{152}$Sm.}
\begin{tabular}{ccc|ccc}
\hline\hline
$I^{\pi}_i$ & $I^{\pi}_f$  &  TQOCH (W.U.)  & $I^{\pi}_i$ & $I^{\pi}_f$  &  TQOCH (W.U.) \\
\hline
$2^-_{\text{band12}}$   &   $2^+_{\text{band1}}$  & 14.1   &  $9^-_{\text{band12}}$   &   $8^+_{\text{band6}}$  &  2.28\\
$2^-_{\text{band12}}$   &   $4^+_{\text{band1}}$  &  5.52  &  $9^-_{\text{band12}}$   &   $9^+_{\text{band3}}$  &  1.22\\
$2^-_{\text{band12}}$   &   $4^+_{\text{band5}}$  &  2.71  &  $9^-_{\text{band12}}$   &   $9^+_{\text{band6}}$  &  2.66\\
$2^-_{\text{band12}}$   &   $4^+_{\text{band6}}$  &  2.75  &  $9^-_{\text{band12}}$   &   $10^+_{\text{band1}}$ &  8.75\\
$2^-_{\text{band12}}$   &   $5^+_{\text{band6}}$  &  9.52  &  $9^-_{\text{band12}}$   &   $10^+_{\text{band3}}$ &  1.34\\
$3^-_{\text{band12}}$   &   $0^+_{\text{band1}}$  &  5.53  &  $9^-_{\text{band12}}$   &   $10^+_{\text{band4}}$ &  1.37\\
$3^-_{\text{band12}}$   &   $4^+_{\text{band1}}$  &  12.5  &  $10^-_{\text{band12}}$  &   $7^+_{\text{band3}}$  &  1.03\\
$3^-_{\text{band12}}$   &   $4^+_{\text{band5}}$  &  1.52  &  $10^-_{\text{band12}}$  &   $8^+_{\text{band1}}$  &  9.01\\
$3^-_{\text{band12}}$   &   $4^+_{\text{band6}}$  &  7.79 &  $10^-_{\text{band12}}$  &   $8^+_{\text{band6}}$  &  1.50\\
$3^-_{\text{band12}}$   &   $6^+_{\text{band1}}$  &  1.32  &  $10^-_{\text{band12}}$  &   $9^+_{\text{band3}}$  &  1.37\\
$3^-_{\text{band12}}$   &   $6^+_{\text{band6}}$  &  8.27  &  $10^-_{\text{band12}}$  &   $9^+_{\text{band6}}$  &  3.87\\
$4^-_{\text{band12}}$   &   $2^+_{\text{band1}}$  &  7.74  &  $10^-_{\text{band12}}$  &   $10^+_{\text{band6}}$ &  1.66\\
$4^-_{\text{band12}}$   &   $4^+_{\text{band1}}$  &  3.78  &  $4^-_{\text{band10}}$   &   $4^+_{\text{band5}}$  & 17.5 \\
$4^-_{\text{band12}}$   &   $4^+_{\text{band5}}$  &  3.82  &  $4^-_{\text{band10}}$   &   $4^+_{\text{band6}}$  & 19.2 \\
$4^-_{\text{band12}}$   &   $4^+_{\text{band6}}$  &  1.08  &  $4^-_{\text{band10}}$   &   $5^+_{\text{band6}}$  & 63.9 \\
$4^-_{\text{band12}}$   &   $5^+_{\text{band6}}$  &  4.02  &  $4^-_{\text{band10}}$   &   $6^+_{\text{band5}}$  & 4.78 \\
$4^-_{\text{band12}}$   &   $6^+_{\text{band1}}$  &  8.01  &  $4^-_{\text{band10}}$   &   $6^+_{\text{band6}}$  & 29.1 \\
$4^-_{\text{band12}}$   &   $6^+_{\text{band6}}$  &  1.77  &  $4^-_{\text{band10}}$   &   $7^+_{\text{band6}}$  & 7.52 \\
$4^-_{\text{band12}}$   &   $7^+_{\text{band6}}$  &  6.10  &  $5^-_{\text{band10}}$   &   $4^+_{\text{band5}}$  & 30.0 \\
$5^-_{\text{band12}}$   &   $2^+_{\text{band1}}$  &  4.89  &  $5^-_{\text{band10}}$   &   $4^+_{\text{band6}}$  & 23.1 \\
$5^-_{\text{band12}}$   &   $4^+_{\text{band1}}$  &  1.61  &  $5^-_{\text{band10}}$   &   $5^+_{\text{band6}}$  & 2.35 \\
$5^-_{\text{band12}}$   &   $4^+_{\text{band5}}$  &  1.88  &  $5^-_{\text{band10}}$   &   $6^+_{\text{band5}}$  & 4.65 \\
$5^-_{\text{band12}}$   &   $5^+_{\text{band6}}$  &  5.94  &  $5^-_{\text{band10}}$   &   $6^+_{\text{band6}}$  & 23.3 \\
$5^-_{\text{band12}}$   &   $6^+_{\text{band1}}$  &  10.6  &  $5^-_{\text{band10}}$   &   $7^+_{\text{band6}}$  & 47.9 \\
$5^-_{\text{band12}}$   &   $6^+_{\text{band6}}$  &  2.60 &  $5^-_{\text{band10}}$   &   $8^+_{\text{band6}}$  & 13.5 \\
$5^-_{\text{band12}}$   &   $7^+_{\text{band6}}$  &  3.28  &  $6^-_{\text{band10}}$   &   $4^+_{\text{band5}}$  & 14.8 \\
$5^-_{\text{band12}}$   &   $8^+_{\text{band1}}$  &  1.67  &  $6^-_{\text{band10}}$   &   $4^+_{\text{band6}}$  & 12.3 \\
$5^-_{\text{band12}}$   &   $8^+_{\text{band6}}$  &  3.97  &  $6^-_{\text{band10}}$   &   $5^+_{\text{band6}}$  & 27.0 \\
$6^-_{\text{band12}}$   &   $4^+_{\text{band1}}$  &  8.97  &  $6^-_{\text{band10}}$   &   $6^+_{\text{band5}}$  & 3.76 \\
$6^-_{\text{band12}}$   &   $5^+_{\text{band6}}$  &  2.74  &  $6^-_{\text{band10}}$   &   $6^+_{\text{band6}}$  & 16.0 \\
$6^-_{\text{band12}}$   &   $6^+_{\text{band1}}$  &  1.72  &  $6^-_{\text{band10}}$   &   $7^+_{\text{band6}}$  & 9.49 \\
$6^-_{\text{band12}}$   &   $6^+_{\text{band5}}$  &  1.43  &  $6^-_{\text{band10}}$   &   $8^+_{\text{band5}}$  & 4.87 \\
$6^-_{\text{band12}}$   &   $6^+_{\text{band6}}$  &  3.94  &  $6^-_{\text{band10}}$   &   $8^+_{\text{band6}}$  & 36.4 \\
$6^-_{\text{band12}}$   &   $8^+_{\text{band1}}$  &  8.38  &  $6^-_{\text{band10}}$   &   $9^+_{\text{band6}}$  & 22.4 \\
$6^-_{\text{band12}}$   &   $8^+_{\text{band6}}$  &  3.85  &  $7^-_{\text{band10}}$   &   $4^+_{\text{band6}}$  & 5.18 \\
$6^-_{\text{band12}}$   &   $9^+_{\text{band6}}$  &  3.61  &  $7^-_{\text{band10}}$   &   $5^+_{\text{band6}}$  & 38.0 \\
$7^-_{\text{band12}}$   &   $4^+_{\text{band1}}$  &  4.01  &  $7^-_{\text{band10}}$   &   $6^+_{\text{band5}}$  & 3.02 \\
$7^-_{\text{band12}}$   &   $6^+_{\text{band1}}$  &  2.91  &  $7^-_{\text{band10}}$   &   $6^+_{\text{band6}}$  & 4.98 \\
$7^-_{\text{band12}}$   &   $6^+_{\text{band3}}$  &  1.36  &  $7^-_{\text{band10}}$   &   $7^+_{\text{band6}}$  & 30.4 \\
$7^-_{\text{band12}}$   &   $6^+_{\text{band5}}$  &  2.40  &  $7^-_{\text{band10}}$   &   $9^+_{\text{band6}}$  & 41.9 \\
$7^-_{\text{band12}}$   &   $6^+_{\text{band6}}$  &  1.53  &  $7^-_{\text{band10}}$   &   $10^+_{\text{band6}}$ & 10.7 \\
$7^-_{\text{band12}}$   &   $7^+_{\text{band6}}$  &  4.40  &  $8^-_{\text{band10}}$   &   $5^+_{\text{band6}}$  & 11.6 \\
$7^-_{\text{band12}}$   &   $8^+_{\text{band1}}$  &  9.44  &  $8^-_{\text{band10}}$   &   $6^+_{\text{band5}}$  & 6.66 \\
$7^-_{\text{band12}}$   &   $8^+_{\text{band5}}$  &  1.31  &  $8^-_{\text{band10}}$   &   $6^+_{\text{band6}}$  & 33.6 \\
$7^-_{\text{band12}}$   &   $9^+_{\text{band3}}$  &  1.19  &  $8^-_{\text{band10}}$   &   $7^+_{\text{band6}}$  & 1.41 \\
$7^-_{\text{band12}}$   &   $9^+_{\text{band6}}$  &  4.01  &  $8^-_{\text{band10}}$   &   $8^+_{\text{band5}}$  & 4.56 \\
$7^-_{\text{band12}}$   &   $10^+_{\text{band1}}$ &  1.62  &  $8^-_{\text{band10}}$   &   $8^+_{\text{band6}}$  & 26.7 \\
$8^-_{\text{band12}}$   &   $6^+_{\text{band1}}$  &  9.16  &  $8^-_{\text{band10}}$   &   $10^+_{\text{band5}}$ & 1.14 \\
$8^-_{\text{band12}}$   &   $6^+_{\text{band6}}$  &  1.28  &  $8^-_{\text{band10}}$   &   $10^+_{\text{band6}}$ & 14.3 \\
$8^-_{\text{band12}}$   &   $7^+_{\text{band6}}$  &  3.98  &  $9^-_{\text{band10}}$   &   $6^+_{\text{band6}}$  & 18.4 \\
$8^-_{\text{band12}}$   &   $8^+_{\text{band6}}$  &  2.93  &  $9^-_{\text{band10}}$   &   $7^+_{\text{band6}}$  & 37.1 \\
$8^-_{\text{band12}}$   &   $10^+_{\text{band1}}$ &  8.29  &  $9^-_{\text{band10}}$   &   $9^+_{\text{band6}}$  & 32.1 \\
$8^-_{\text{band12}}$   &   $10^+_{\text{band6}}$ &  2.98  &  $10^-_{\text{band10}}$  &   $7^+_{\text{band6}}$  & 23.1 \\
$9^-_{\text{band12}}$   &   $6^+_{\text{band1}}$  &  3.15  &  $10^-_{\text{band10}}$  &   $8^+_{\text{band5}}$  & 4.32 \\
$9^-_{\text{band12}}$   &   $6^+_{\text{band4}}$  &  1.11 &  $10^-_{\text{band10}}$  &   $8^+_{\text{band6}}$  & 26.6 \\
$9^-_{\text{band12}}$   &   $8^+_{\text{band1}}$  &  3.76  &  $10^-_{\text{band10}}$  &   $9^+_{\text{band6}}$  & 1.29 \\
$9^-_{\text{band12}}$   &   $8^+_{\text{band3}}$  &  2.67  &  $10^-_{\text{band10}}$  &   $10^+_{\text{band5}}$ & 1.35 \\
$9^-_{\text{band12}}$   &   $8^+_{\text{band4}}$  &  1.20  &  $10^-_{\text{band10}}$  &   $10^+_{\text{band6}}$ & 13.4 \\
$9^-_{\text{band12}}$   &   $8^+_{\text{band5}}$  &  1.59 \\                                                             \hline
\hline
\end{tabular}
\end{center}
\end{table}

\clearpage

\bibliography{reference}

\end{document}